\documentclass[preprint, superscriptaddress, showpacs,preprintnumbers,amsmath,amssymb,endfloats*]{revtex4}
\usepackage{graphicx}

\begin{document}

\thispagestyle{empty}

\title{Lateral Casimir force between sinusoidally
corrugated surfaces: Asymmetric profiles, deviations from the
proximity force approximation and comparison with
exact theory}

\author{H.-C.~Chiu}
\affiliation{Department of Physics and Astronomy, University of
California, Riverside, California 92521, USA}

\author{G.~L.~Klimchitskaya}
\email{Galina.Klimchitskaya@itp.uni-leipzig.de}
\affiliation{North-West Technical University, Millionnaya St. 5,
St.Petersburg, 191065, Russia}

\author{V.~N.~Marachevsky}
\email{maraval@mail.ru}
\affiliation{V.A.~Fock Institute of Physics,
Saint-Petersburg State University, St.Petersburg, 198504, Russia}

\author{V.~M.~Mostepanenko}
\email{Vladimir.Mostepanenko@itp.uni-leipzig.de}
\affiliation{Noncommercial Partnership ``Scientific Instruments'',
Tverskaya St. 11, Moscow, 103905, Russia}

\author{U.~Mohideen}
\email{Umar.Mohideen@ucr.edu}
\affiliation{Department of Physics and
Astronomy, University of California, Riverside, California 92521,
USA}



\begin{abstract}
The  lateral Casimir force, which arises between aligned sinusoidally
corrugated surfaces of a sphere and a plate, was measured for the case of
a small corrugation period beyond the applicability region of the
proximity force approximation. The increased amplitudes of the corrugations
on both the sphere and the plate allowed observation of an asymmetry of the
lateral Casimir force, i.e., deviation of its profile from a perfect sine
function. The dependences of the lateral force on the phase shift between
the corrugations on both test bodies were measured at different separations
in two sets of measurements with different amplitudes of corrugations
on the sphere. The maximum magnitude of the lateral force as a function of
separation was also measured in two successive experiments. All measurement
data were compared with the theoretical approach using the proximity
force approximation and with the exact theory based on Rayleigh expansions
with no fitting parameters.
In both cases real material properties of the test bodies and nonzero
temperature were taken into account. The data were found to be in a good
agreement with the exact theory but deviate significantly from the
predictions of the proximity force approximation
 approach. This provides the quantitative confirmation
for the observation of diffraction-type effects that are disregarded within
the PFA approach. Possible applications of the phenomenon of the lateral
Casimir force in nanotechnology for the operation of micromachines
are discussed.
\end{abstract}
\maketitle

\section{Introduction}

The Casimir effect \cite{1} is presently well known due to the
many potential applications in both fundamental
physics and nanotechnology. The Casimir force is an extension of the
van der Waals interaction to larger separations between
macroscopic bodies where the relativistic retardation of the
electromagnetic interaction becomes important due to the finite
speed of light. The Casimir effect originates from the existence
of zero-point and thermal oscillations in restricted quantization
volumes. The results of extensive experimental
and theoretical studies on the role
of the Casimir force in configurations with idealized boundary surfaces
(infinitely thin, with ideal-metal boundary conditions etc.) and also
between surfaces of real material bodies with account of roughness,
nonzero skin depth and nonzero temperature are presented in Ref.~\cite{2}
(see also earlier books and reviews \cite{3,4,5,6,7,8,9,10}).
Interrelationship between
experiment and theory in the case of real materials
was specially reviewed in Ref.~\cite{11}. In condensed matter physics
the Casimir force is taken into account in the investigation of various
properties of thin films, surface tension, in atomic force microscopy
and critical phenomena \cite{12}. When the characteristic sizes of
microdevices shrink below a micrometer, the role of the Casimir and
electric forces become comparable. This opens new opportunities for the
creation of nanoscale devices actuated by the Casimir force \cite{13,14}.
Investigation of the combined action of the electric and Casimir forces in
microdevices is vital for the resolution of issues
associated with their
stability \cite{15,16}.

The most universally known {\it normal} Casimir force acts in the
direction perpendicular to the interacting surfaces. However, when the
 material properties of the interacting bodies are anisotropic or
they are asymmetrically positioned,
 a {\it lateral} Casimir force may exist which
acts tangential to the surface. Similar to the normal Casimir force, the
lateral force originates from the modification of electromagnetic
zero-point and thermal oscillations by material boundaries. For plates
made of anisotropic materials, the lateral Casimir force and related torques
were theoretically predicted in Refs.~\cite{17,18} (see also
Refs.~\cite{4,19}). For two parallel ideal metal plates covered with uniaxial
sinusoidal corrugations of equal periods, the lateral Casimir force was
predicted\cite{20,21} and calculated\cite{22,23} in the
second perturbation order with respect to the amplitudes of corrugations.
In both cases a sinusoidal
dependence of the result on the respective characteristic
angle was obtained.
For sinusoidally corrugated plates made of real metal described by the
plasma model, the lateral Casimir force was considered in Refs.~\cite{24,25}
in second order perturbation theory. In the case of two ideal metal
parallel plates covered with periodic uniaxial corrugations of rectangular
shape, some results for the normal Casimir force were found in
Refs.~\cite{26,27}. It is pertinent to note that the use of normal and
lateral Casimir forces gives the possibility to actuate both normal and
lateral translations of corrugated surfaces in micromachines by means
of the electromagnetic zero-point oscillations. The lateral Casimir
force might be used to solve the tribological problems plaguing the
microdevice industry. Specifically, it was suggested \cite{28,29,30}
to use this effect for frictionless transmission of lateral motion.
There are also proposals \cite{31,32} to measure the Casimir torque
arising between the anisotropic bodies due to the lateral Casimir force.

Both the normal and lateral Casimir forces are in general nonadditive and
possess a complicated dependence on the geometrical shape of the
boundary surfaces connected with diffraction effects. In the configuration
of a smooth Al coated sphere above a sinusoidally corrugated Al coated
plate the nontrivial behavior of the normal Casimir force was
experimentally demonstrated \cite{33}. This was done in an
additive regime when the sphere-plate separation is much smaller than
the period of corrugations \cite{23}. Recently \cite{34} the nontrivial
behavior of the normal Casimir force caused by the complicated geometry
of a boundary surface was observed in the configuration of a smooth Au
coated sphere above a Si plate covered with rectangular corrugations
(trenches). In this case the nonadditive regime was used where separation
distances are of the order of corrugation period. The deviations of the
measurement data from the additive theory were reported \cite{34}.
These deviations are, however, about 50\% less than those expected for
ideal metal boundaries in the theory taking exact account of the
geometrical shape \cite{27}. The remaining discrepancy might be explained
by the role of a nonzero skin depth which was not accounted for in
Ref.~\cite{27}. Thus, the scattering approach for
grating geometries taking the exact account
of surface geometry at zero temperature and describing metallic properties
by means of the simple plasma model \cite{35,36} was found to be in
better agreement with the data of Ref.~\cite{34}.

The first measurement of the lateral Casimir force was performed between the
surfaces of an Au coated sphere and an Au coated plate both covered with
aligned uniaxial corrugations of sinusoidal shape \cite{37,38}.
The corrugation amplitudes were $A_1=59\,$nm on the plate and $A_2=8\,$nm
on the sphere. The period of corrugations was equal to 1200\,nm, i.e.,
much larger than separation distances between the sphere and plate.
This means that the experiments of Refs.~\cite{37,38} was done in the
application region of the proximity force approximation (PFA) \cite{2,8,11}.
The lateral Casimir force  with an amplitude of $3.2\times 10^{-13}\,$N
at the shortest separation of 221\,nm was found to sinusoidally
oscillate as a function of the phase shift between the corrugations.
The total experimental error of the force
amplitude measurement at the closest
point was determined to be around 19\% at the 95\% confidence
level \cite{2}.
The experimental results were compared with theory using the PFA which
should supposedly be applicable at separations much smaller than the
corrugation period. In so doing the dielectric properties of Au were
described using the simple plasma model in fourth order perturbation
theory with respect to a small parameter related to the ratio
of the skin depth to the separation distance. The experimental results were
found to be in good agreement with the theory in the limits of the
experimental error. It was also predicted theoretically (but not observed
experimentally due to the use of small corrugation amplitudes and
insufficient precision) that the lateral Casimir force is asymmetric,
i.e., deviates from purely sinusoidal harmonic dependence on the phase
shift, through the contribution of higher order terms in the
perturbation expansion with respect to the corrugation amplitudes.

In this paper we present complete measurement data of the recently performed
experiment on the lateral Casimir force between a sinusoidally
corrugated  sphere and  plate covered with Au layers. The much smaller
period of corrugations used in this experiment provides opportunities to
measure deviations of the lateral Casimir force from the theoretical
predictions obtained on the basis of the PFA. Much deeper corrugations
on both the sphere and the plate and shorter separations where the
measurements
were performed make it possible to demonstrate the asymmetry of the lateral
Casimir force, as predicted in Ref.~\cite{38}. The measurement data are
compared with the exact theory describing the Rayleigh scattering of the
electromagnetic oscillations on the sinusoidally corrugated boundary
surfaces of a sphere and a plate with no fitting
parameters. Only the configuration of a perfect
sphere near a plane plate is treated using the PFA. At the separations
considered, this leads to only a negligibly small error of order
0.1\% \cite{2,11}. The dielectric properties of Au are described by means
of the generalized plasma-like model \cite{39,40} which takes into account
the interband transitions of core electrons. The computational results
for the amplitude of the lateral Casimir force as a function of separation
and for the lateral Casimir force as a function of the phase shift at
different separations obtained using the exact theory at the
laboratory temperature
($T=300\,$K) are found to be in a very good agreement with
the measurement data. The same data deviate markedly from the theoretical
results computed using the PFA. This clearly demonstrates the influence
of diffraction effects on the lateral Casimir force.

Some of the results of this paper related to the first set of our
measurements were briefly published in Ref.~\cite{41}. Here, we present
additional phase dependences of the lateral Casimir force not included
in Ref.~\cite{41} for the same corrugation amplitudes, as were used in
\cite{41}, and also the results of further experiments performed
with a
much larger corrugation amplitude on the sphere. We demonstrate that the
experimental data of various measurements for different samples are in
mutual agreement. We also present additional important evidence
characterizing the quality of corrugated surfaces, procedures of
electrostatic calibrations, and details of error analysis. Calculations
using the PFA are done by means of the exact Lifshitz formula rather than
with the help of perturbation theory in corrugation amplitudes, as in
Ref.~\cite{41}. The role of surface roughness is estimated. The details of
exact computations of the lateral Casimir force using the Rayleigh
scattering theory  at nonzero temperature given in this paper
have never been presented previously.

The paper is organized as follows. In Sec.\ II we describe the
experimental setup  and sample preparation. Section\ III contains the
description of the procedures of electrostatic calibration using both the
normal and lateral electric forces. In Sec.\ IV the measurement data
and the error analysis are presented. Section\ V is devoted to the
computation of the lateral Casimir force between corrugated surfaces
using the PFA. In Sec.\ VI the lateral Casimir force is computed
in a more fundamental way using the Rayleigh theory. In both Secs.\ V
and VI the theoretical results are compared with the experimental data.
Section\ VII contains our conclusions and discussion.

\section{Experimental setup and sample preparation}

A schematic diagram of the experimental setup is shown in Fig.~\ref{setup}.
The experiments under discussion are performed using a standard atomic force
microscope (AFM) in oil-free vacuum chamber at a pressure below 10\,mTorr
 and at room temperature.
The observation of the lateral Casimir force is done using two test bodies
whose surfaces are covered with longitudinal sinusoidal corrugations of
the same period. The axes of the corrugations should be perfectly parallel.
Following \cite{38}, one can see that misalignment by $1^{\circ}$ of the 
axes of
corrugations can lead to the loss of any lateral force.
As the first test body, we have used a sinusoidally corrugated grating
of size $5\times 5\,\mbox{mm}^2$ vertically mounted on the piezotube of
an AFM (see Fig.~\ref{setup}). Here, the corrugations have an average period
$\Lambda=574.7\,$nm (i.e., more than two times smaller than in
Refs.~\cite{37,38} in order to achieve the regime where the PFA becomes
inappropriate for the calculation of the lateral Casimir force between
the corrugated surfaces). The amplitude of corrugations on the grating
$A_1=85.4\pm 0.3\,$nm was  a factor of 1.45 larger than in
Refs.~\cite{37,38}. A $320\,\mu$m long V-shaped silicon nitride
cantilever of the AFM was first uniformly coated with 40\,nm of Al.
This modification of cantilever was performed to improve its thermal and
electric conductivity and prevent deformation due to differential thermal
expansion in vacuum.

A polystyrene sphere of a $200\pm 4\,\mu$m diameter was placed at the end of
the cantilever with conductive Ag epoxy. Next, a freshly cleaved mica sheet of
$400\,\mu$m length, $200\,\mu$m width, and a few micrometer thickness was
attached to the bottom of the sphere also with Ag epoxy. Then a second
polystyrene sphere with a nominal diameter $2R=200\pm 4\,\mu$m was
attached to the bottom free end of the mica sheet (see Fig.~\ref{setup}).
After imprint of corrugations on its surface (see below), this last sphere
was used as the second test body in the measurement of the lateral
Casimir force. We emphasize that the silver epoxy is rigid at all attachments.
The first sphere and mica sheet are needed to isolate the
laser reflection spot on the cantilever tip from the interaction region
between the two corrugated surfaces. As a result, the effect of scattered
light from the top and sides of the corrugated plate is
substantially reduced. The resulting system (cantilever, mica sheet and
the attached two spheres) was uniformly coated with a 10\,nm layer of Cr
and then with a more than 50\,nm layer of Au in a thermal evaporator.
Uniformity of coating was ensured through rotation
of the cantilever \cite{42}. The Cr layer was used
to improve the adhesion of the Au coating and prevent its peeling during
the imprint of the corrugations on the sphere.

The lateral Casimir force arises when the uniaxial corrugations on the two
test bodies are perfectly aligned and have the same period $\Lambda$
\cite{20,21,22,23,37,38}. To ensure that these conditions are satisfied,
we have imprinted corrugations on the second sphere using the grating as
a template. In order to obtain deeper corrugations than in Refs.~\cite{37,38},
a more sophisticated procedure was implemented. Here, the template grating had
a 300\,nm Au coating applied by the manufacturer on top of sinusoidal
corrugations made of hard epoxy on a 3-mm-thick Pyrex substrate. Note that
hard epoxy does not require the use of the
oxide of Al layer which was coated
on the soft plastic grating in Refs.~\cite{37,38}. The smaller period of
corrugations used here required the use of a precise stepper motor and
piezocontrolled imprinting technique. The imprinting procedure was done as
follows. At first, the second sphere was brought in contact with the
grating using a micromanipulator. Then a hard flat stylus
(see Fig.~\ref{setup}) was moved in $10\,\mu$m steps using a stepper motor
until it comes into contact with the other side of the sphere which
is now
sandwiched against the grating. Next the imprinting was done by applying
a voltage to the $z$ piezo to gently press the sphere between the grating
and the hard flat stylus. To obtain deeper imprints while preserving
sphericity some additional pressure was put on the sphere using the
stepper motor. Special care was needed at this step. If the applied pressure
from the hard flat stylus was too much, the resulting rotation of the
second sphere will cause the cantilever to break. There is less concern on the
preservation of sphericity as polystyrene spheres are elastic.
Then the voltage on the $z$-piezo is reversed to remove the
pressure on the sphere.
With the help
of the same stepper motor the hard flat stylus
 was gently removed and the sphere was
then translated horizontally
by around $200\,\mu$m to a different position on the grating.
The last operation is necessary because the part of the grating used to
imprint corrugations on the sphere might have changed its local amplitude
during the imprinting process. This translation was performed preserving
the orientation of the cantilever.
The same imprinting procedure, but with
application of larger pressure by means of the stepper motor, was done with
a similar sphere in the second set of our measurements. After the imprint
process was done, the system was left in vacuum for
more than 2 hours to reach a
stable equilibrium.

The corrugations on the grating and on both spheres were examined using an
AFM and found to be rather homogeneous. This was done after the completion
of the experiment. An AFM scan of the Au-coated corrugations of the
vertical template
grating is shown in Fig.~\ref{rghPl}(a). The value of the amplitude of these
corrugations $A_1$, also measured using the AFM, is indicated above.
The variance of the stochastic roughness on the grating was estimated in
the following way. The topography of the groove from an AFM scan
[see Fig.~\ref{rghPl}(a)] was fitted to a sine function.
Note that the top and bottom of corrugations are symmetric as
expected for a sine. This is illustrated
in Fig.~\ref{rghPl}(b)
where a typical section of Fig.~\ref{rghPl}(a) along a
$y=\rm{const}$ plane is shown.
Then the difference
of heights between the groove topography and the sine function
was calculated at a sufficient number of intermediate points.
The variance was found using these differences as the root-mean-square
deviation between the two curves. A total 30 grooves
were used to evaluate the average value of the variance leading to
$\delta_1=13\,$nm. In Fig.~\ref{rghSph}(a) we show an AFM scan of the
corrugations imprinted on the sphere used in the first set of our
measurements. The amplitude of these corrugations over a
$30\times 30\,\mu\mbox{m}^2$ area on the sphere was measured with an AFM
to be $A_2=13.7\pm 0.4\,$nm (i.e., an increase by a factor of 1.7 as
compared with Refs.~\cite{37,38}).
Figure~\ref{rghSph}(b) is a typical section of Fig.~\ref{rghSph}(a)
along a $y=\rm{const}$ plane. It illustrates a symmetry between
the top and bottom of the imprinted corrugations.
The variance of the stochastic roughness
of the imprinted grooves $\delta_2$
was found using the same procedure as for a grating
with the result $\delta_2=2.7\,$nm.

In Fig.~\ref{sph1} the AFM scan of a  $5\times 5\,\mu\mbox{m}^2$
area
of imprinted corrugations on the sphere used in the first set of
measurements is demonstrated. Here, the lighter tone shows higher points,
as indicated in the additional scale. As is seen in Fig.~\ref{sph1},
the imprinting procedure did not destroy the sphericity of the imprinted
surface. An AFM scan of the sphere surface with imprinted corrugations
used in the second set of our measurements was also performed.
The amplitude of these corrugations was measured to be
$\tilde{A}_2=25.5\pm 0.6\,$nm (increase by a factor of 3.2 as
compared with Refs.~\cite{37,38}). The respective variance of the
stochastic roughness in this case was
$\tilde{\delta}_2=8.8\,$nm. Note that
the relatively large values of the variance
of the stochastic roughness presented above
(in comparison with the case of flat
surfaces \cite{43}) are explained by the fact that the r.m.s. deviation
is calculated along the $z$-axis, i.e., not perpendicular to the
corrugated surface at most of the intermediate points. In Fig.~\ref{sph2}
we show a $5\times 5\,\mu\mbox{m}^2$ area of the corrugations imprinted on
the sphere used in the second set of our measurements.
As can be seen in this
figure, the grooves are deeper than for the sphere used in the first
set of measurements, and the sphericity of the surface is preserved.
The diameters of the spheres with the imprinted
corrugations used in the first and
second set of measurements were measured to be
$2R=194.0\pm 0.3\,\mu$m and $2\tilde{R}=194.8\pm 0.3\,\mu$m, respectively,
using a scanning electron microscope (SEM). The SEM was calibrated
with a NIST-traceable calibration grating which was
independently cross calibrated with our interferometrically
calibrated AFM.

The distinctive feature of our setup is that the lateral Casimir force
acting along the $x$-axis tangential to the corrugated sphere and a
grating leads to the vertical bending of the cantilever. This bending is
measured using bicell photodiodes A and B  in Fig.~\ref{setup}.
Whereas a force acting normal to the test bodies (the normal Casimir
force) leads to a torsional deflection of the cantilever. The torsional
spring constant of the used cantilever $K_{\rm tor}$ was found to be 46
times larger than the bending spring constant $K_{\rm ben}$ (see the
next section).
Because of this the normal Casimir force could lead to only a
negligible change in the position of the second sphere and in the phase
of the corrugations. Thus the setup used is much more sensitive
to detecting the lateral Casimir force, while simultaneously suppressing
the effect of the normal Casimir force.

\section{Electrostatic calibrations}

The calibration of the cantilever (i.e., the determination of its spring
constants), the measurement of the residual potential difference $V_0$
between the sphere and the plate, and the determination of  the
separation on contact $z_0$ were done by using  both the normal and
lateral electric forces. We begin with the determination of $K_{\rm tor}$
and $V_0$ by means of the normal electric force. For this purpose we
have measured the torsional cantilever deflection due to the voltages
$V$ applied to the grating while the sphere remained grounded.
This was performed through the measurement of the difference
signal between the bicell photodiodes shown in Fig.~\ref{setup}.
To determine the parameters of our interest, the measured deflection
signal $S_{\rm nor}^{\rm el}$ at some separation $a$ between the zero
levels of corrugations on the grating and on the sphere, where the Casimir
force is negligible, should be fitted to the theoretical expression for
the normal electrostatic force between the corrugated surfaces.
To attain these ends we begin with a brief derivation of such an expression
in the sphere-plate configuration which is not available in the literature.

    In the configuration of an ideal metal sphere above an ideal metal
plate with smooth boundary surfaces the exact expression for the
electric force is well known \cite{44}. In the range of separations
from $a=100\,$nm to $a=6\,\mu$m it can be represented \cite{45} by the
following polynomial with a relative error less than $10^{-4}$\%
\begin{eqnarray}
&&
F_{\rm nor}^{\rm el, sp}(a)=-2\pi\epsilon_0(V-V_0)^2\Phi(a),
\nonumber \\
&&
\Phi(a)=\sum\limits_{i=-1}^{6}c_i\left(\frac{a}{R}\right)^i,
\label{eq1}
\end{eqnarray}
\noindent
where $\epsilon_0$ is the permittivity of free space and the
numerical coefficients $c_i$ are given by
\begin{eqnarray}
&&
c_{-1}=0.5,\quad c_0=-1.18260,\quad c_1=22.2375, \quad
c_2=-571.366,
\nonumber \\
&&
c_3=9592.45 , \quad c_4=-90200.5,\quad c_5=383084.,
\quad c_6=-300357.
\label{eq2}
\end{eqnarray}
\noindent
The approximate expression for the electric force in the case when the
surfaces of a sphere and a plate are covered with sinusoidal
corrugations can be obtained using the PFA, as applied to electric forces.
For this purpose we consider the separation along the $z$-axis
between any two points $z_s$ on the corrugated sphere and $z_p$
on the corrugated plate
\begin{equation}
z_s-z_p=a+A_2\sin(2\pi x/\Lambda +\varphi)-A_1\sin(2\pi x/\Lambda).
\label{eq3}
\end{equation}
\noindent
Here $a$, as defined above, is the closest separation between a perfectly
shaped sphere and a plate, and $\varphi\equiv 2\pi x_0/\Lambda$ is
the phase shift between the corrugations of both bodies. Then, according to
the PFA, the normal electric force between corrugated a sphere and a plate
is obtained from Eq.~(\ref{eq1}) by averaging over a period as
\begin{equation}
F_{\rm nor}^{\rm el}(a,\varphi)=-2\pi\epsilon_0(V-V_0)^2
\frac{1}{\Lambda}\int_{0}^{\Lambda}dx\,\Phi(z_s-z_p).
\label{eq4}
\end{equation}
\noindent
The integral in (\ref{eq4}) can be calculated by using the following
substitution of (\ref{eq3}):
\begin{equation}
z_s-z_p=a\left[1+\beta\cos(2\pi x/\Lambda-\alpha)\right],
\label{eq5}
\end{equation}
\noindent
where
\begin{eqnarray}
&&
\beta\equiv\beta(a,\varphi)=\frac{1}{a}(A_1^2+A_2^2-
2A_1A_2\cos\varphi)^{1/2},
\nonumber \\
&&
\tan\alpha=(A_2\cos\varphi-A_1)/(A_1\sin\varphi).
\label{eq6}
\end{eqnarray}
\noindent
Substituting Eq.~(\ref{eq5}) into Eq.~(\ref{eq4}) and performing integration,
one obtains
\begin{eqnarray}
&&
F_{\rm nor}^{\rm el}(a,\varphi)=-2\pi\epsilon_0(V-V_0)^2
\left[\frac{R}{2a}\,\frac{1}{\sqrt{1-\beta^2}}\right.
\nonumber \\
&&
~~~
+c_0+c_1\frac{a}{R}+c_2\frac{a^2(2+\beta^2)}{2R^2}+
c_3\frac{a^3(2+3\beta^2)}{2R^3}
\nonumber \\
&&
~~~
+c_4\frac{a^4(8+24\beta^2+3\beta^4)}{8R^4}
+c_5\frac{a^5(8+40\beta^2+15\beta^4)}{8R^5}
\nonumber \\
&&
~~~\left.
+c_6\frac{a^6(16+120\beta^2+90\beta^4+5\beta^6)}{16R^6}\right].
\label{eq7}
\end{eqnarray}

It is well known that for corrugated surfaces the use of the PFA may not
lead to precise expressions for the Casimir force \cite{23,24,25}.
For a static electric force, however, the PFA should work much better because
in this case the diffraction effects are absent. To verify the validity of
Eq.~(\ref{eq7}), we applied a 3-dimensional finite element analysis
(FEA) using Comsol Multiphysics \cite{46} to numerically solve the Poisson
equation with appropriate boundary conditions in the configuration of
corrugated sphere and plate.  In the FEA computation, it is difficult
to use the whole corrugated sphere and plate due to computer limitations.
This necessitates the use of truncated spheres and parts of the plates.
Such a replacement is justified as the primary contribution to the force
comes from sphere-plate regions which are in close proximity. However,
the error introduced by the truncation has to be independently confirmed
to be negligible by varying the size of the regions of the sphere and
plate
used in the computations.

Computations were performed as a function of the common size $l$ of a
truncated section of the corrugated sphere and a square corrugated plate.
The separation distance and phase difference between the corrugations on
both surfaces were set to $a=250\,$nm and $\varphi=\pi/2$, respectively.
The system was enclosed in a grounded rectangular box which represents
the boundary conditions at infinity. Next, boundary conditions (applied
voltages) identical to the experimental parameters were assigned to all
objects. The rectangular enclosure and the sphere were both set to be
grounded. The rectangular box was automatically set to be at infinity by
Comsol Multiphysics. Next the Poisson equation was solved for the given
boundary conditions. Then the normal electrostatic force was calculated by
integrating the $zz$-component of the Maxwell stress tensor over the
surface of corrugated sphere and compared with Eq.~(\ref{eq7}).
To make sure that the solution converged, the size of the corrugated
objects and the number of the surface mesh elements were both varied
and the force recalculated. For each $l$, the number of the mesh elements
was increased from one million to 22 millions until the calculated force
converged. The size of the truncated section of the corrugated plate and
sphere was increased from $l=14\,\mu$m to $l=50\,\mu$m.
The convergence of the force was observed for $l>45\,\mu$m.
The corresponding difference of Eq.~(\ref{eq7}) from the FEA results varied
from --36\% to 2.8\% when $l$ increased from $14\,\mu$m to $50\,\mu$m,
respectively, with convergence at the largest values of $l$. Thus, we
confirmed from simulations 
that Eq.~(\ref{eq7}) is valid with an error less
than 2.8\%.
It turns out that this error is almost phase independent and, thus,
does not propagate to the lateral electric force.

Now we return to the experimental electrostatic calibrations using the
normal electric force. The deflection signal $S_{\rm nor}^{\rm el}$ was
measured for eight different voltages between --0.52\,V and 0.47\,V applied
to the grating at the constant grating-sphere separation $a=1\,\mu$m
where the Casimir force is negligibly small
(see below for the determination of absolute separations using the lateral
electrostatic force). The respective experimental normal electrostatic
force is given by
\begin{equation}
F_{\rm nor}^{\rm el}(a,\varphi)=k_{\rm tor}S_{\rm nor}^{\rm el}(a,\varphi),
\label{eq8}
\end{equation}
\noindent
where $k_{\rm tor}$ is the normal force calibration constant measured in the
units of force per unit deflection signal (note that for the attractive
force the deflection signal is negative). This constant is connected with
the torsional spring constant discussed above   as
\begin{equation}
k_{\rm tor}=K_{\rm tor}m_{\rm tor},
\label{eq9}
\end{equation}
\noindent
where the deflection coefficient $m_{\rm tor}$ is measured in the units
of length per unit deflection signal. The obtained experimental data for
$S_{\rm nor}^{\rm el}$ and for the respective force $F_{\rm nor}^{\rm el}$
from Eq.~(\ref{eq8}) was fitted to Eq.~(\ref{eq7}). The resulting mean
values of $V_0$ and $k_{\rm tor}$ found from the fit are
\begin{equation}
V_0=-39.6\pm 1.6\,\mbox{mV}, \quad
k_{\rm tor}=7.00\pm 0.08\,\mbox{nN/unit}{\ } S.
\label{eq10}
\end{equation}

Measurement of the lateral Casimir force as a function of absolute separation
requires knowledge of the lateral force calibration constant
\begin{equation}
k_{\rm ben}=K_{\rm ben}m_{\rm ben},
\label{eq11}
\end{equation}
\noindent
where $m_{\rm ben}$ is the bending deflection coefficient.
If the separation on contact $z_0$ is determined,
absolute separation between the mean values of the corrugations on both
surfaces is given by \cite{4,11}
\begin{equation}
a=z_0+z_{\rm piezo}+S_{\rm nor}m_{\rm tor},
\label{eq12}
\end{equation}
\noindent
where $z_{\rm piezo}$ is the distance moved by the plate owing to the
voltage applied to the piezoelectric actuator, $S_{\rm nor}$ is the
photodiode difference signal due to the force (either electric or Casimir).

The determination of $k_{\rm ben}$ and $z_0$ is achieved by measuring the
cantilever deflection signal due to the lateral electrostatic force which
arises when a voltage is applied to the grating. The measurements of
this signal were performed at small separations from close to $z_0$ to
$z_0+120\,$nm. Note that calibrations using the lateral electrostatic force
were done after the measurements of the deflection signal due to the
lateral Casimir force are performed, but are reported in this section for
the benefit of the reader. First, a voltage of 141.456\,mV was applied to the
grating. The sphere was kept at a distance 3.96\,nm from $z_0$. The phase
shift between corrugations was changed continuously at a frequency 0.103\,Hz
with the $x$-piezo to a maximum translation of $3.3\,\mu$m.

It is important that corrugated surfaces move parallel to each other as
the phase is changed with the $x$-piezo. The grating template was mounted as
parallel as possible to the $x$-axis on the AFM piezo. However,
experimentally there is always a finite nonzero angle between the grating
template and the $x$-axis. When the phase change is introduced by moving
the $x$-piezo, this tilt will lead to changes in the separation distance
between the corrugated surfaces, which in turn would result in a systematic
error of decreasing or increasing force with phase. This nonparallelity
should be corrected before start of the experiments using the lateral
electrostatic force. A voltage is applied to the plate and the lateral
electrostatic force is monitored as a function of the phase over many periods
of the grating. A correcting voltage adapted from that applied to the
$x$-piezo using an adjustable voltage divider was synchronously applied
to the $z$-piezo. The applied voltage to
the $z$-piezo was changed till the
amplitude of the lateral electrostatic force become independent of the phase.
As an example, in Fig.~\ref{elFc} we show the deflection signal due to
the lateral electrostatic force versus the phase difference $x_0$ between
the corrugated surfaces (a) before and (b) after the tilt correction was
introduced. This was done approximately 25\,nm away from contact
with a
0.2\,V applied voltage. It can be seen that the amplitudes of each peak are
more identical in Fig.~\ref{elFc}(b). This procedure experimentally
verifies that the separation distance between the corrugations remains
fixed as the phase was changed with the $x$-piezo.

The cantilever deflection signal $S_{\rm lat}^{\rm tot}$ corresponding
to the total force acting at the separation $a=z_0+3.96\,$nm
(sum of the lateral electrostatic and lateral Casimir forces) was
recorded at 8192 evenly spaced data points. The sphere was moved further
away from the grating by 5.40\,nm to a separation 9.36\,nm from $z_0$,
and the measurement was repeated. The cantilever deflection signal
$S_{\rm lat}^{\rm tot}$ was measured at many other $a$ at the same voltage
and also for a second voltage of 101.202\,mV applied to the grating.
To obtain the deflection signal due to the lateral electrostatic force
alone, it is necessary to subtract from the obtained signal
$S_{\rm lat}^{\rm tot}$ the deflection signal $S_{\rm lat}^{\rm C}$
due to the lateral Casimir force (the signal $S_{\rm lat}^{\rm C}$ was
measured before $S_{\rm lat}^{\rm tot}$ and the respective measurement
data are reported in Sec.~IV). To perform this subtraction,
$S_{\rm lat}^{\rm C}$ was fitted to the following sum of harmonics:
\begin{equation}
S_{\rm lat}^{\rm C}(a,\varphi)=\sum\limits_{k=1}^{5}A_k(a)\sin(k\varphi),
\label{eq13}
\end{equation}
\noindent
which takes proper account of the fact that the lateral Casimir force
is asymmetric (see Sec.~IV). As the sphere-grating separations
in the measurements of the lateral Casimir force and of the total force
are not identical, interpolation was used to determine the values of
$A_k(a)$ at the separations corresponding to the total lateral force.
After determination of the coefficients $A_k(a)$ from the fit, the signal
$S_{\rm lat}^{\rm C}$ was subtracted from the data for the total
deflection signal $S_{\rm lat}^{\rm tot}$. The obtained deflection
signal $S_{\rm lat}^{\rm el}$ now corresponds to the lateral electrostatic
force
\begin{equation}
F_{\rm lat}^{\rm el}(a,\varphi)=k_{\rm ben}
S_{\rm lat}^{\rm el}(a,\varphi).
\label{eq14}
\end{equation}
\noindent
The experimental force data in this equation were used to determine the
calibration parameter $k_{\rm ben}$ and $z_0$ from the fit to the
theoretical expression for the lateral electric force.

To derive the analytic expression for the lateral electric force acting
between the sinusoidally corrugated surfaces of a sphere and a plate,
we start with the electrostatic energy in sphere-plane configuration
with smooth surfaces \cite{47,48}
\begin{equation}
E^{\rm el,sp}(a)=-2\pi\epsilon_0(V-V_0)^2R\left[c_{-1}\ln\frac{R}{a}+
\tilde{c}-\sum\limits_{i=0}^{6}\frac{c_i}{i+1}
\left(\frac{a}{R}\right)^{i+1}\right].
\label{eq15}
\end{equation}
\noindent
Here, the integration constant $\tilde{c}$ is equal to \cite{47}
\begin{equation}
\tilde{c}=\frac{1}{2}\ln{2}+\frac{23}{40}+\frac{\theta}{126},
\label{eq16}
\end{equation}
\noindent
where $0\leq\theta\leq 1$, and $c_i$ are defined in Eq.~(\ref{eq2}).

Using the PFA, we obtain the electrostatic energy between sinusoidally
corrugated surfaces of a sphere and a plate in the form
\begin{equation}
E^{\rm el}(a,\varphi)=\frac{1}{\Lambda}\int_{0}^{\Lambda}dx
E^{\rm el,sp}(z_s-z_p),
\label{eq17}
\end{equation}
\noindent
where $z_s-z_p$ is given by Eqs.~(\ref{eq3}), (\ref{eq5}) and (\ref{eq6}).
Using the definition of the lateral electric force
[see the formula for the phase shift below Eq.~(\ref{eq3})]
\begin{equation}
F_{\rm lat}^{\rm el}(a,\varphi)=
-\frac{\partial}{\partial x_0}E^{\rm el}(a,\varphi)=
-\frac{2\pi}{\Lambda}\,\frac{\partial}{\partial\varphi}E^{\rm el}(a,\varphi)
\label{eq18}
\end{equation}
\noindent
and calculating all integrals in (\ref{eq17}), we arrive at the result
\begin{eqnarray}
F_{\rm lat}^{\rm el}(a,\varphi)&=&\pi^2\epsilon_0(V-V_0)^2R
\frac{A_1A_2}{a^2\Lambda}\left[
\vphantom{\sum\limits_{i=1}^{6}}
\frac{2}{\sqrt{1-\beta^2}(1+\sqrt{1-\beta^2})}\right.
\nonumber \\
&-&\left.
2\sum\limits_{i=1}^{6}ic_i\left(\frac{a}{R}\right)^{i+1}
Y_i\right].
\label{eq19}
\end{eqnarray}
\noindent
Here, the parameter $\beta\equiv\beta(a,\varphi)$ is defined in
Eq.~(\ref{eq6}) and the coefficients $Y_i\equiv Y_i(\beta^2)$ are
given by
\begin{eqnarray}
&&
Y_1=Y_2=1,\quad Y_3=1+\frac{1}{4}\beta^2,\quad
Y_4=1+\frac{3}{4}\beta^2,
\nonumber \\
&&
Y_5=1+\frac{4}{3}\beta^2+\frac{1}{8}\beta^4, \quad
Y_6=1+10\beta^2+\frac{5}{8}\beta^4.
\label{eq20}
\end{eqnarray}

By fitting the force data in Eq.~(\ref{eq14}) to Eq.~(\ref{eq19}),
the quantities $k_{\rm ben}$ and $z_0$ were found. This was repeated
for four different electrostatic force measurements and the average
values obtained are
\begin{equation}
z_0=117.3\pm 3.0\,\mbox{nm},\quad
k_{\rm ben}=1.27\pm 0.06\,\mbox{nN/unit}{\ }S.
\label{eq21}
\end{equation}
\noindent
Using an independent quad-cell AFM measurement of the torsional movement of
such a cantilever, the torsional spring constant was found to be
$K_{\rm tor}=0.28\pm 0.5\,$N/m. The value of the bending deflection
coefficient $m_{\rm ben}=209\pm 3\,$nm/unit $S$ was obtained in the same
way as in normal Casimir force measurements \cite{42,49} from the change in
position of the contact point for application of different voltages to the
plate during the normal electrostatic force measurement. Then the value of
the bending spring constant $K_{\rm ben}=(6.1\pm 0.3)\times 10^{-3}\,$N/m
was found from Eqs.~(\ref{eq11}) and (\ref{eq21}). By comparing the two
spring constants, we find that $K_{\rm tor}/K_{\rm ben}\approx 46$
which means that our system is really much more sensitive to the lateral
rather than to the normal Casimir force.

In recent literature \cite{50} there is a discussion
 that in several measurements
of the Casimir force the residual potential difference $V_0$ depends on
separation, whereas in some other it is constant. Because of this, we have
performed an additional control measurement of $V_0$ at short
separation $a=127.3\,$nm using the parabolic dependence of
$F_{\rm lat}^{\rm el}$ in Eq.~(\ref{eq19}) on $V$. For this purpose five
different voltages close to the residual potential difference were
applied to the grating and the cantilever deflection was measured as a
function of V leading to $V_0=-39.4\,$mV. This is consistent with the
result (\ref{eq10}) obtained at large separation $a=1\,\mu$m and confirms
that in our experiment the residual potential difference is separation
independent.

The same calibration procedures, as described above for the first set of
measurements, were repeated for a second set with a larger corrugation
amplitude of 25.5\,nm on the sphere.
Here we obtained a residual potential
difference $\tilde{V}_0=-28.35\pm 1.15\,$mV and the following values
of the separation on contact and of the lateral force calibration
constant
\begin{equation}
\tilde{z}_0=131.4\pm 3.8\,\mbox{nm},\quad
\tilde{k}_{\rm ben}=2.05\pm 0.11\,\mbox{nN/unit}{\ }S.
\label{eq21a}
\end{equation}
\noindent
Typically in our second set of measurements
with the larger amplitude corrugations the calibration errors
are a bit larger than in the first, where smaller amplitude
corrugations were used.

\section{Measurement data for the lateral Casimir force and error
analysis}

Now we describe how the measurements of the lateral Casimir force were
performed.  In this case the residual voltage $V_0$,
as determined in the electrostatic calibration using the normal electric force,
was applied to the grating in order to make the electric force
equal to zero. The $x$-piezo was used to move the grating
along the $x$-axis and thus change $\varphi$.
The $z$-piezo, which was independently controlled by an external voltage
source, was used to change $a$.
The piezo extensions with applied voltage in both directions were
calibrated using optical interferometry \cite{51}.

Initially the corrugated sphere was positioned
3.79\,nm from the separation on
contact between the two surfaces $z_0$ determined by the corrugations and
the highest roughness peaks. The thermomechanical drift of the separation
distance was measured to be 0.14\,nm/min from the difference in the
$z$-piezo voltage to bring about contact of the two corrugated surfaces
after a time interval around 30 minutes.
A phase shift was introduced by moving the $x$-piezo
continuously for a total distance of $3.3\,\mu$m at 0.103\,Hz.
The photodiode signal corresponding to the cantilever deflection was
filtered with a low-pass filter with a 30\,ms time constant and
recorded at each of the 8192 points corresponding to $x$-changes of
0.4\,nm.  The effect of the scattered laser light which would lead to a
linear modification of the signal with phase was found to be negligible
in this experiment.

Then the separation from $z_0$ was
increased by 3.6\,nm from 3.79\,nm to 7.39\,nm and the deflection signal
$S_{\rm lat}^{\rm C}$  was similarly measured as a function of
$\varphi$ and recorded. After this, the separation from
$z_0$ was increased by
3.96\,nm from 7.39\,nm to 11.35\,nm and the measurements
repeated. Next $S_{\rm lat}^{\rm C}$ due to
the lateral Casimir force as a function of
$\varphi$ was measured at separations of 20.05, 32.48,
45.30, 58.01 and 70.86\,nm from  $z_0$.

Using the values of $k_{\rm ben}$ and $z_0$ obtained through the
electrostatic calibration (see Sec.~III) the measurement data for the
deflection signal $S_{\rm lat}^{\rm C}$ were converted into values of
the lateral Casimir force at every separation $a$ as:
\begin{equation}
F_{\rm lat}^{\rm C}(a,\varphi)=k_{\rm ben}S_{\rm lat}^{\rm C}(a,\varphi).
\label{eq22}
\end{equation}
\noindent
The resulting lateral Casimir force for a sphere with the corrugation
amplitude $A_2=13.7\,$nm over four corrugation periods is shown in
Fig.~\ref{phaseN}(a-c) as dots versus the normalized
lateral displacement
$x/\Lambda$ between the corrugated surfaces at separations
$a=124.7$, 128.6, and 149.8\,nm, respectively. Similar results were
obtained at all other $a$ listed above. As can be seen in Fig.~\ref{phaseN},
the lateral Casimir force is a periodic function of the phase shift.
The maximum values of the lateral force, $f\equiv\max|F_{\rm lat}^{\rm C}|$,
decrease with the increase of $a$. The important characteristic feature of the
periodic curves in Fig.~\ref{phaseN}(a-c) is that they are
{\it asymmetric}, i.e., the dependence of $F_{\rm lat}^{\rm C}$ on
$x/\Lambda$ is not strictly sinusoidal (note that the sinusoidal
dependence of the lateral force on the phase shift holds only if the
calculation is restricted to the lowest order in corrugation amplitudes
$\sim A_1A_2$; see Sec.~V), The true asymmetry of the measured lateral
force is obvious even without the theory curve. For example, in
Fig.~\ref{phaseN}(a) the average shift of each maximum point from the
midpoint of two adjacent minima is $(0.12\pm 0.2)\Lambda$.
With the increase of separation, the relative contribution of higher
perturbation orders in a small parameter $A_1A_2/a^2$ decreases and the
dependence of the lateral force on the phase shift becomes more close
to sinusoidal. The solid lines in Fig.~\ref{phaseN} are related to the
exact theoretical computations. They are discussed in Sec.~VI.

The experimental values of the $\max|F_{\rm lat}^{\rm C}|$ versus
separation $a$ are shown in Fig.~\ref{ampN} as crosses. The arms of the
crosses indicate the total experimental errors determined at a 95\%
confidence level using the following procedure.
{}From the data of the phase curves in Fig.~\ref{phaseN} (and related
curves at all the other separations measured) we have calculated the mean
values $\bar{f}$ and the variances $s_{\bar{f}}$ of the quantity
$f=\max|F_{\rm lat}^{\rm C}|$ at each $a$ (see columns 2 and 3 in Table 1).
For the first set of measurements under consideration, the
averaging was performed over the four periods. The respective Student
coefficient is $t_{(1+\gamma)/2}(3)=3.18$ where $\gamma=0.95$.
Thus, the experimental random error is obtained from the data in column 3
of Table 1 as
\begin{equation}
\Delta^{\!\rm rand}f(a)=s_{\bar{f}}t_{0.975}(3).
\label{eq23}
\end{equation}
\noindent
The main sources of systematic errors in the quantity $f$
are the errors due to
the uncertainty in $k_{\rm ben}$ and due to the resolution of data.
According to Eq.~(\ref{eq21}), $\Delta k_{\rm ben}/k_{\rm ben}=0.047$.
From Eq.~(\ref{eq22}) this leads to the systematic error
$\Delta_1^{\!\rm syst}f(a)=0.047f(a)$. The systematic error resulting from the
resolution of data does not depend on separation
$\Delta_2^{\!\rm syst}f(a)=0.035\,$pN. Combining these two errors at a 95\%
confidence level as random quantities characterized by a uniform
distribution \cite{2,52,53}, we obtain the values of the total systematic
error given in column 4 of Table 1. Using the rule for the combination
of random and systematic errors at the same confidence level
\cite{2,52,53}, we obtain the total experimental error $\Delta^{\!\rm tot}f$
as a function of separation presented in the fifth column of Table 1.
As can be seen in Table 1, all absolute errors are larger at shorter
separations and decrease with increasing separation. This might be
explained by local deviations of the shape of corrugations from perfect
sinusoidal form which are more influential at shorter separations.
The total relative error of $\max|F_{\rm lat}^{\rm C}|$ at $a=121.1\,$nm is
equal to 22.5\% (at a 95\% confidence level). The relative errors at
separations $a=124.7$, 128.6, and 137.3\,nm are equal to 13\%, 25.5\%,
and 14\%, respectively.

The error in the measurement of the absolute separations, $\Delta a$,
is a combination of the error in the determination of $z_0$,
$\Delta z_0=3\,$nm, indicated in Eq.~(\ref{eq21}) and the error in the
quantity $z_{\rm piezo}+S_{\rm nor}m_{\rm tor}$ in Eq.~(\ref{eq12}).
The latter is equal to the half of the first step of the $z$-piezo,
$\Delta_pz\approx 2\,$nm. Combining these two errors at a 95\%
confidence level with the help of the same rule, as applied above for
the systematic errors \cite{2,52,53}, one obtains $\Delta a=4\,$nm.
This is a factor of eight improvement as compared with Ref.~\cite{38}.
Note that the bands between the dashed and the solid lines
in Fig.~\ref{ampN} are related
to theoretical predictions using the PFA and the exact theory,
respectively. They are discussed in Secs.~V and VI.

The same measurement procedure, as described above, was applied in the
second set of measurements for a sphere covered with deeper
corrugations of an amplitude $\tilde{A}_2$. The resulting lateral Casimir
force over five or six corrugation periods is shown in
Fig.~\ref{phaseM}(a-c) as dots versus the lateral
displacement at separations
$a=134$, 156.5 and 179\,nm, respectively. Similar results were obtained
also at separations $a=145.2$, 201.6, 224, and 246.6\,nm.
In the same way, as in Fig.~\ref{phaseN}, the lateral Casimir force at
each $a$ is a periodic function which is asymmetric
(nonsinusoidal). The asymmetry is more pronounced at
shorter separations. The experimental values of the
$\max|F_{\rm lat}^{\rm C}|$ versus separation are plotted in Fig.~\ref{ampM}
as crosses. Similar to Fig.~\ref{ampN} the arms of the crosses show the
total experimental errors of force and separation measurements determined
at a 95\% confidence level. Equation (\ref{eq21a}) and the same statistical
procedure, as in the first set of measurements, were used to determine
these errors with the only difference that for the second set the averaging
was performed at all separations over five periods [the respective
Student coefficient is $t_{(1+\gamma)/2}(4)=2.78$ with $\gamma=0.95$].
The total error in the measurement of separation, $\Delta a=4.7\,$nm,
turned out to be a bit larger than in the first set of measurements.
The total error of the lateral force measurements demonstrates similar
irregular behavior on separation distance. This  can also be explained
by the influence of local deviations of groove shape
from sinusoidal form keeping in
mind that with larger corrugation amplitude the role
of such deviations should be larger.
In Table 2 we present the mean values, the variances of the mean,
the systematic errors, and the total experimental errors
(at a 95\% confidence level) of the maximum magnitudes of the
lateral Casimir force at different separations for the second set
of measurements.
The total relative error of the lateral force measurements at separations
$a=134$, 145.2, 156.5, and 179\,nm varies as 23\%, 16\%, 14\%, and 13\%,
respectively, all calculated at a 95\% confidence level.

\section{Computation of the lateral Casimir force using the proximity
force approximation}

In this section and in Sec.~VI we compare the obtained experimental
results for the lateral Casimir force with two theoretical approaches
applicable in the case of corrugated surfaces: the PFA and the exact
scattering approach, respectively, with no
fitting parameters. As was mentioned in the Introduction,
the PFA approach was also used to compare with theory the measurement
data of the first observation of the lateral Casimir force in
Refs.~\cite{37,38}. However, in those papers
the real properties of Au were
described using the simple plasma model and fourth-order perturbation
theory with respect to the relative skin depth at zero temperature.
Here, we develop a more complete description of the experimental data
in the framework of the PFA approach based on the Lifshitz formula
at the laboratory temperature. This will help us to separate diffraction type
contributions to the lateral Casimir force which are beyond the PFA.

We start with the Lifshitz formula for the
normal Casimir force acting between
a sphere and a plate made of a real material (Au) but bounded with perfectly
smooth surfaces \cite{2,8,11}
\begin{eqnarray}
F^{\rm C,sp}(a)&=&k_BTR\sum_{l=0}^{\infty}
{\vphantom{\sum}}^{\prime}\int_{0}^{\infty}
k_{\bot}\,dk_{\bot}\left\{
\ln\left[1-r_{\rm TM}^2(i\xi_l,k_{\bot})\,e^{-2q_la}\right]\right.
\nonumber \\
&+&\left.
\ln\left[1-r_{\rm TE}^2(i\xi_l,k_{\bot})\,e^{-2q_la}\right]\right\}.
\label{eq24}
\end{eqnarray}
Here, $k_B$ is the Boltzmann constant, $\xi_l=2\pi k_BTl/\hbar$ with
$l=0,\,1,\,2,\,\ldots$ are the Matsubara frequencies, prime near the
summation sign adds a multiple 1/2 to the term with $l=0$.
The reflection coefficients for the transverse magnetic and transverse
electric polarizations of the electromagnetic field are defined as
\begin{equation}
r_{\rm TM}(i\xi_l,k_{\bot})=
\frac{\varepsilon_lq_l-k_l}{\varepsilon_lq_l+k_l},\quad
r_{\rm TE}(i\xi_l,k_{\bot})=\frac{q_l-k_l}{q_l+k_l},
\label{eq25}
\end{equation}
\noindent
where
\begin{equation}
q_l^2=k_{\bot}^2+\frac{\xi_l^2}{c^2}, \quad
k_l^2=k_{\bot}^2+\varepsilon_l\frac{\xi_l^2}{c^2}
\label{eq26}
\end{equation}
\noindent
and $\varepsilon_l\equiv\varepsilon(i\xi_l)$ is the dielectric permittivity
for the material of the sphere and plate calculated along the imaginary
Matsubara frequencies. Equation (\ref{eq24}) is obtained from the
standard Lifshitz expression for the free energy in the configuration
of two parallel plates by the multiplication by $2\pi R$ in accordance with
the PFA. Keeping in mind the values of the experimental parameters
($R\sim  100\,\mu$m, $a\sim 100\,$nm), one can conclude that the error
introduced by the use of the PFA in this case (of about
$a/R=0.1$\% \cite{8,54,55,56}) is negligibly small.

Now we consider the sphere and the plate covered with sinusoidal corrugations
described in Secs.~II and III. Within the PFA approach, the approximate
expression for the normal Casimir force acting between corrugated surfaces
of a sphere and a plate can be obtained from Eq.~(\ref{eq24}) by replacing
$a$ with $z_s-z_p$ defined in Eq.~(\ref{eq3}) and averaging over the period
of corrugations $\Lambda$. By expanding also the logarithms in Eq.~(\ref{eq24})
into a series, we arrive at
\begin{eqnarray}
&&F_{\rm nor}^{\rm C}(a,\varphi)=-\frac{k_BTR}{\Lambda}
\sum\limits_{n=1}^{\infty}\frac{1}{n}\sum_{l=0}^{\infty}
{\vphantom{\sum}}^{\prime}\int_{0}^{\infty}
k_{\bot}\,dk_{\bot}\nonumber \\
&&
~~~\times\left[r_{\rm TM}^{2n}(i\xi_l,k_{\bot})+
r_{\rm TE}^{2n}(i\xi_l,k_{\bot})\right]\,e^{-2q_lna}
\label{eq27} \\
&&
~~~\times\int_{0}^{\Lambda}dx\,
e^{-2q_ln[A_2\sin(2\pi x/\Lambda+\varphi)-A_1\sin(2\pi x/\Lambda)]}.
\nonumber
\end{eqnarray}
\noindent
{}From this expression it is simple to obtain the Casimir energy in the
configuration of a sphere above a plate both covered with sinusoidal
corrugations
\begin{eqnarray}
&&
E^{\rm C}(a,\varphi)=\int_a^{\infty}\!\!dzF_{\rm nor}^{\rm C}(z,\varphi)
\nonumber \\
&&~~
=-\frac{k_BTR}{2\Lambda}
\sum\limits_{n=1}^{\infty}\frac{1}{n^2}\sum_{l=0}^{\infty}
{\vphantom{\sum}}^{\prime}\int_{0}^{\infty}
\frac{k_{\bot}\,dk_{\bot}}{q_l}\nonumber \\
&&
~~~\times\left[r_{\rm TM}^{2n}(i\xi_l,k_{\bot})+
r_{\rm TE}^{2n}(i\xi_l,k_{\bot})\right]\,e^{-2q_lna}
\label{eq28} \\
&&
~~~\times\int_{0}^{\Lambda}dx\,
e^{-2q_ln[A_2\sin(2\pi x/\Lambda+\varphi)-A_1\sin(2\pi x/\Lambda)]}.
\nonumber
\end{eqnarray}
\noindent
Introducing the dimensionless variables
\begin{equation}y=2aq_l, \qquad \zeta_l=\frac{2a\xi_l}{c}
\label{eq29}
\end{equation}
\noindent
and performing the integration with respect to $x$, one can rearrange
Eqs.~(\ref{eq27}) and (\ref{eq28}) to the form
\begin{eqnarray}
&&
F_{\rm nor}^{\rm C}(a,\varphi)=-\frac{k_BTR}{4a^2}
\sum\limits_{n=1}^{\infty}\frac{1}{n}\sum_{l=0}^{\infty}
{\vphantom{\sum}}^{\prime}\int_{\zeta_l}^{\infty}
y\,dy\nonumber \\
&&
~~~\times\left[r_{\rm TM}^{2n}(i\zeta_l,y)+
r_{\rm TE}^{2n}(i\zeta_l,y)\right]\,e^{-ny}\,{I}_0(n\beta y),
\label{eq30} \\
&&
E^{\rm C}(a,\varphi)=-\frac{k_BTR}{4a}
\sum\limits_{n=1}^{\infty}\frac{1}{n^2}\sum_{l=0}^{\infty}
{\vphantom{\sum}}^{\prime}\int_{\zeta_l}^{\infty}
dy\nonumber \\
&&
~~~\times\left[r_{\rm TM}^{2n}(i\zeta_l,y)+
r_{\rm TE}^{2n}(i\zeta_l,y)\right]\,e^{-ny}\,{I}_0(n\beta y).
\label{eq31}
\end{eqnarray}
\noindent
Here, $I_0(z)$ is the Bessel function of an imaginary argument and
$\beta$ is defined in Eq.~(\ref{eq6}).

The general expression for the lateral Casimir force can be obtained from
Eq.~(\ref{eq31}) in the same way as the lateral electric force in
Eq.~(\ref{eq18}). After performing
the differentiation, we rearrange the result
to the form
\begin{eqnarray}
&&
F_{\rm lat}^{\rm C}(a,\varphi)=\frac{\pi k_BTRA_1A_2}{2a^3\Lambda\beta}
\sin\varphi
\sum\limits_{n=1}^{\infty}\frac{1}{n}\sum_{l=0}^{\infty}
{\vphantom{\sum}}^{\prime}\int_{\zeta_l}^{\infty}
y\,dy\nonumber \\
&&
~~~\times\left[r_{\rm TM}^{2n}(i\zeta_l,y)+
r_{\rm TE}^{2n}(i\zeta_l,y)\right]\,e^{-ny}\,{I}_1(n\beta y).
\label{eq32}
\end{eqnarray}
\noindent
This expression is convenient for numerical computations.

Equations (\ref{eq30}) and (\ref{eq32}) are generalizations
for the case of real materials of the previously known
expressions obtained at $T=0$ for ideal metal corrugated bodies.
In this case $r_{\rm TM}^2=r_{\rm TE}^2=1$ and one obtains
\begin{eqnarray}
&&
F_{\rm nor}^{\rm C}(a,\varphi)=-\frac{\hbar cR}{8\pi a^3}
\sum\limits_{n=1}^{\infty}\frac{1}{n}\int_{0}^{\infty}
y^2\,dy\,e^{-ny}I_0(n\beta y),
\label{eq33} \\
&&
F_{\rm lat}^{\rm C}(a,\varphi)=\frac{\hbar cRA_1A_2}{4 a^4\Lambda\beta}
\sin\varphi
\sum\limits_{n=1}^{\infty}\frac{1}{n}\int_{0}^{\infty}
y^2\,dy\,e^{-ny}I_1(n\beta y).
\nonumber
\end{eqnarray}
\noindent
Introducing the new variable $v=ny$ and performing the summation,
we arrive at
\begin{eqnarray}
&&
F_{\rm nor}^{\rm C}(a,\varphi)=-\frac{\pi^3\hbar cR}{720a^3}
\int_{0}^{\infty}dv\,v^2\,e^{-v}I_0(\beta v),
\label{eq34} \\
&&
F_{\rm lat}^{\rm C}(a,\varphi)=\frac{\pi^4\hbar cRA_1A_2}{360a^4\Lambda\beta}
\sin\varphi
\int_{0}^{\infty}dv\,v^2\,e^{-v}I_1(\beta v).
\nonumber
\end{eqnarray}
\noindent
After performing the integration in these equations, the final result is
\begin{eqnarray}
&&
F_{\rm nor}^{\rm C}(a,\varphi)=-\frac{\pi^3\hbar cR}{720a^3}\,
\frac{2+\beta^2}{(1-\beta^2)^{5/2}},
\label{eq35} \\
&&
F_{\rm lat}^{\rm C}(a,\varphi)=
\frac{\pi^4\hbar cRA_1A_2}{360a^4\Lambda(1-\beta^2)^{5/2}}
\sin\varphi.
\nonumber
\end{eqnarray}
\noindent
The last expression for the lateral force was obtained in Refs.~\cite{37,38}
with corrections due to the skin depth. In the first perturbation order
in $A_1A_2$ (i.e., for $\beta=0$) this expression was also used in
Refs.~\cite{24,25}.

Equations (\ref{eq30}) and (\ref{eq32}) take into account the sphericity
of the second test body and sinusoidal corrugations on both bodies in
the framework of the PFA and the nonzero skin depth in the framework
of the Lifshitz theory.
To perform numerical computations, one should use a model
to represent the dielectric permittivity.
At the separations considered (from 100 to 250\,nm)
the thermal effects do not play any role in any
of the theoretical approaches and
reliable results can be obtained using the generalized plasma-like
model \cite{2,39,40}
\begin{equation}
\varepsilon(i\xi_l)=1+\frac{\omega_p^2}{\xi_l^2}+
\sum\limits_{j=1}^{6}\frac{g_j}{\omega_j^2+\xi_l^2+\gamma_j\xi_l},
\label{eq36}
\end{equation}
\noindent
where $\omega_p=9.0\,$eV for Au is the plasma frequency. The values of the
oscillator frequencies $\omega_j$, relaxation parameters $\gamma_j$
and the oscillator strengths $g_j$ leading to approximately the same results
for ${\rm Im}\varepsilon(\omega)$ as those
obtained from the optical tabulated data \cite{57} over the frequency
range from 2 to 25\,eV are listed in Refs.~\cite{2,40}.

An important factor that is not taken into account in Eq.~(\ref{eq32})
is the surface roughness. Stochastic roughness covering the corrugated
surfaces in both the sphere and the plate is clearly seen in
Figs.~\ref{rghPl},\,\ref{rghSph} and its measured variances are presented
in Sec.~II. In view of its stochastic (nonperiodic) character, surface
roughness cannot influence the phase dependence of the Casimir energy in the
configuration of corrugated surfaces. However, it contributes to the
Casimir energy, and, thus, to the lateral Casimir force, through a
phase independent correction factor which depends only on separation.
It has been found \cite{2,8} that the correction factor to the Casimir
force due to stochastic roughness in the configuration of a
sphere above a plate with no corrugations is given by
\begin{equation}
\eta_{F}(a)=1+6\frac{\delta_1^2+\delta_2^2}{a^2}+
45\frac{(\delta_1^2+\delta_2^2)^2}{a^4}.
\label{eq37}
\end{equation}
\noindent
Now we take into account that in the framework of the multiplicative
approach the correction factor due to surface
roughness is the same for real and
ideal metals.
Keeping in mind that the Casimir force between a perfectly shaped ideal metal
sphere and a plate is $F(a)\sim 1/a^3$, we obtain the correction to the
Casimir energy by integrating the Casimir force with
the inclusion of roughness,
$\eta_F(a)F(a)$, with respect to $a$
\begin{equation}
\eta_{E}(a)=1+3\frac{\delta_1^2+\delta_2^2}{a^2}+
15\frac{(\delta_1^2+\delta_2^2)^2}{a^4}.
\label{eq38}
\end{equation}
\noindent
Note that for the experimental variances (see Sec.~II) the fourth order
term in Eq.~(\ref{eq38}) is negligibly small. Thus, for the first
set
of measurements at the shortest separation $a=120\,$nm the second order
term contributes 3\% of the force, but the fourth order term only 0.2\%
of the force. For the second set of measurement with deeper
corrugations on the sphere the contributions of the second and fourth order
terms at separation $a=134\,$nm are 4\% and 0.3\%, respectively.

To obtain the correction factor to the lateral Casimir force due
to the
surface roughness with account of the sinusoidal corrugations on the sphere
and on the plate, we should replace $a$ with separation $z_s-z_p$ between
the corrugated surfaces in Eqs.~(\ref{eq3}) and (\ref{eq5}) and perform
the averaging with respect to the phase shift and the period
\begin{eqnarray}
&&
\eta_{\rm corr}(a)=1+3(\delta_1^2+\delta_2^2)
\left\langle\frac{1}{(z_s-z_p)^2}\right\rangle,
\label{eq39} \\
&&
\left\langle\frac{1}{(z_s-z_p)^2}\right\rangle=
\frac{1}{2\pi\Lambda a^2}\int_{0}^{2\pi}d\varphi\int_{0}^{\Lambda}dx
\frac{1}{[1+\beta\cos(2\pi x/\Lambda-\alpha)]^2}.
\nonumber
\end{eqnarray}
\noindent
Calculating both integrals in the second equality in Eq.~(\ref{eq39})
and substituting the obtained result into the first equality, we arrive at
\begin{eqnarray}
&&
\eta_{\rm corr}(a)=1+3
\frac{\delta_1^2+\delta_2^2}{a(a+A_1-A_2)^{1/2}(a+A_2-A_1)^{1/2}}
\nonumber \\
&&
~~~~\times
F\left(\frac{1}{2},\frac{1}{2};1;\frac{4A_1A_2}{(a+A_1-A_2)(a+A_2-A_1)}\right),
\label{eq40}
\end{eqnarray}
\noindent
where $F(\alpha,\beta;\gamma;z)$ is the hypergeometric function.

Substituting the values of $A_i$ and $\delta_i$ from the first set
of measurements (see Sec.~II) in Eq.~(\ref{eq40}), we find that at
separations $a=120$, 150, and 200\,nm the correction factor due to surface
roughness takes the values $\eta_{\rm corr}=1.054$, 1.029, and 1.015,
respectively.
For the second set of measurements (i.e., using $\tilde{A}_2$ and
$\tilde{\delta}_2$ in Sec.~II) the correction factor in  Eq.~(\ref{eq40})
takes the values $\tilde{\eta}_{\rm corr}=1.060$, 1.041, and 1.021 at
separations $a=130$, 150, and 200\,nm, respectively.
It is seen that even for  the second set of our
measurements, where the surface roughness on the sphere is larger than in
the first one, the roughness correction is rather moderate and does not
exceed 6\% of the lateral Casimir force.

In Fig.~\ref{ampN} (the lower dashed line) we present the computational
results for the maximum values of the lateral Casimir force as a function
of separation in the framework of the PFA [i.e., using Eqs.~(\ref{eq32})
and (\ref{eq36})] for the first set of our measurements. This line does not
take surface roughness into account. The upper dashed line in
Fig.~\ref{ampN} presents the results for
$\max|F_{\rm lat}^{\rm C}|\eta_{\rm corr}$ as a function of separation,
i.e., taking into account the correction due to surface roughness.
Keeping in mind that the multiplicative approach provides the means to
estimate the effect of roughness rather than to calculate it precisely
on the basis of the fundamental theory, we consider the band between the
dashed lines in Fig.~\ref{ampN} as a theoretical prediction given by
the PFA approach. As can be seen in Fig.~\ref{ampN}, the experimental data
indicated as crosses are inconsistent with the prediction of the PFA
approach, thus revealing the role of diffraction-type effects on the
corrugations of relatively small period used in this experiment.

In Fig.~\ref{ampM} a similar comparison between the experimental data and
theory using the PFA approach is done for the second set of our
measurements with corrugations of larger amplitude.
The lower dashed line is computed using Eqs.~(\ref{eq32})
and (\ref{eq36}) with the surface roughness disregarded. The upper dashed
line is obtained by the multiplication of the computation results by an
additional factor $\tilde{\eta}_{\rm corr}$ computed
using  Eq.~(\ref{eq40}). Again
the theoretical band between the dashed lines is inconsistent
with the experimental data shown as crosses due to the neglect of
diffraction-like effects.
In the next section we
present a more fundamental theoretical approach taking the diffraction
effects into account which brings theory into agreement with the
experimental data.

\section{Computation of the lateral Casimir force using the
Rayleigh theory}

The PFA approach used to calculate the lateral Casimir force in the
configuration of a sphere above a plate with corrugated boundaries has
an uncontrolled error which increases with decreasing period of
corrugations. As was mentioned in the Introduction, in such cases a more
fundamental theory is desirable which takes exact
account of diffraction effects.
Here, we present such a theory, based on the scattering approach, for
the configuration of two plates covered with periodical corrugations of
arbitrary shape but common period. This theory can be applied to our
experimental configuration by using the PFA for the transition from the
configuration of two plane plates to a smooth sphere above a plate.
According to our discussion in Sec.~V, the error in computations
introduced by the use of the PFA for this restricted purpose alone, is
extremely small and quite satisfactory for the needs of our experiment.

The scattering approach started with the work \cite{58} where it was
applied to obtain the Lifshitz formula and Refs.~\cite{59,60} on the
multi-scattering expansion. On this basis
the multipole scattering
technique was developed in Refs.~\cite{61,62,63} and was applied to
several configurations with curved boundaries for scalar and
electromagnetic cases.
Alternative techniques within the scattering
approach to the Casimir effect were presented in Refs.~\cite{64,65,66,66a}.
For objects of spherical and cylindrical shapes the multipole scattering
technique was used in Refs.~\cite{54,55,67,68,69}.
This technique works well for
large separations between the objects and shows poor convergence
when the separations between the objects are small. At small
separations the leading asymptotic expansion of the Casimir energy
at zero temperature within the scattering approach was developed in
Ref.~\cite{70}.  The
technique developed in Refs.~\cite{35,36} uses another basis, the Rayleigh
basis, which is a generalization of the plane wave basis. The
Rayleigh basis is a natural choice for grating geometries.

We consider two parallel $3$-dimensional longitudinal (along the $y$-axis)
periodic (along the $x$-axis) dielectric (metallic) gratings of
arbitrary form separated by a vacuum gap so that they form a
waveguide and one grating is located above the other (we assume
that the edge of the corrugation region of each grating is
perpendicular to the $z$ axis). The periods of both gratings $\Lambda$
are equal. In general, the top and bottom gratings can have
different shapes and different dielectric (metallic) properties. In
Fig.~\ref{grat1} the height of the corrugation region is equal to $h$
for the bottom grating. The meaning of the  lateral displacement $x_0$
becomes obvious from the comparison of Fig.~\ref{grat1} with
Fig.~\ref{grat2} and Fig.~\ref{grat3}. We suppose that the space
between the two gratings is vacuum with $\varepsilon=\mu=1$ and
we assume $\mu=1$ inside the medium.

The physical problem is time and $y$ invariant, so
that the particular solutions for the electric and
magnetic fields can be written in the form:
\begin{eqnarray}
E_i (x,y,z,t) &=& E_i (x,z) \exp(ik_y y - i \omega t)  ,
\nonumber\\
H_i (x,y,z,t) &=& H_i (x,z) \exp(ik_y y - i \omega t) .
\label{41}
\end{eqnarray}
\noindent
The solutions of Maxwell equations should satisfy the
quasi-periodicity conditions:
\begin{eqnarray}
E_i(x+\Lambda,z)& =& e^{i k_x \Lambda} E_i(x,z) ,
\label{eq42} \\
H_i(x+\Lambda,z) &= &e^{i k_x \Lambda} H_i(x,z) .
\nonumber
\end{eqnarray}
\noindent
In every Casimir problem one needs to determine the complete basis
of solutions. Let us suppose that the top grating is absent. We
consider a generalized conical diffraction problem $(k_y \ne 0)$ on
the lower grating for the incident wave with the $x$-component
of the wave vector equal to some fixed value which we denote
$\tilde{k}_x$.
The longitudinal components of the
electromagnetic field outside the corrugated region ($z \ge h$) may
be written by making use of  the Rayleigh
expansion \cite{71} for an incident monochromatic wave:
 \begin{eqnarray}
E_y(x,z) &=&
I_{\tilde{k}_x}^{(e)} \exp (i \tilde{k}_x x - i \tilde{\beta}^{(1)}z)
\nonumber \\
 &&+\sum_{n=-\infty}^{\infty}
R_{np}^{(e)} \exp( i \alpha_n x  + i \beta_n^{(1)} z)  ,
\label{eq43} \\
H_y(x,z) &=&
I_{\tilde{k}_x}^{(h)} \exp (i \tilde{k}_x x - i \tilde{\beta}^{(1)}z)
\nonumber \\
 &&+\sum_{n=-\infty}^{\infty} R_{np}^{(h)} \exp( i
\alpha_n x  + i \beta_n^{(1)} z).
\nonumber
\end{eqnarray}
\noindent
Here
\begin{equation}
\alpha_n = k_x + 2\pi n/ \Lambda
\label{eq44a}
\end{equation}
\noindent
with
\begin{equation}
k_x=\tilde{k}_x-\frac{2\pi}{\Lambda}
\left[\frac{\Lambda\tilde{k}_x}{2\pi}\right]
\equiv\tilde{k}_x-\frac{2\pi}{\Lambda}p ,
\label{eq44b}
\end{equation}
\noindent
where $[r]$ is the integer part of the number $r$.
{}From this it follows that
$0\leq k_x<2\pi/\Lambda$.
The other notation in Eq.~(\ref{eq43}) is as follows:
\begin{equation}
{\tilde{\beta}^{(1)}}
{\vphantom{\tilde{\beta}^{(1)}}}^2 = \omega^2
- k_y^2 - \tilde{k}_x^2,\quad
{\beta_n^{(1)}}
{\vphantom{\beta_n^{(1)}}}^2 = \omega^2
- k_y^2 - \alpha_n^2 ,
\label{eq44c}
\end{equation}
\noindent
The quantities $I_{\tilde{k}_x}^{(e,h)}$ and  $R_{np}^{(e,h)}$
in Eq.~(\ref{eq43}) are the Rayleigh
coefficients. [Note that with notations (\ref{eq44a})--(\ref{eq44c}) it holds
$\tilde{k}_x\equiv\alpha_p$,
$\tilde{\beta}^{(1)}\equiv\beta_p^{(1)}$.]
The solution (\ref{eq43}) is also valid outside any periodic structure in $x$
direction. All other
field components can be expressed in terms of the longitudinal
components $E_y, H_y$ by using standard formulas as done
in waveguide theory. This can be done since the factor $\exp(i
k_y y)$ is conserved after the reflection of the electromagnetic
wave from the grating.

 At $z=0$
the solution has to satisfy the expansions
 \begin{eqnarray}
E_y(x,z) &=&  \sum_{n=-\infty}^{\infty}
T_{np}^{(e)} \exp( i \alpha_n x  - i \beta_n^{(2)} z)  ,
\label{eq45} \\
H_y(x,z) &=&  \sum_{n=-\infty}^{\infty}
T_{np}^{(h)} \exp( i \alpha_n x  - i \beta_n^{(2)} z),
\nonumber
 \end{eqnarray}
where
\begin{equation}
{\beta_n^{(2)}}^{ 2} = \varepsilon \omega^2 - k_y^2 - \alpha_n^2  ,
\label{eq46}
\end{equation}
\noindent
which are valid for $z\leq 0$. The coefficients $T_{np}^{(e,h)}$
are called the transmission matrix coefficients.

The $E_y$ component of the electromagnetic field in the region $0\le
z \le h$ is defined as follows:
\begin{equation}
E_y(x,z) = \sum_{n=-\infty}^{\infty} E_y^n (z) \exp(i\alpha_n x) .
\label{eq49}
\end{equation}
\noindent
The other components of the electromagnetic field in the region $0\le z
\le h$ are defined in analogy.
Inside the corrugation
region $0 \le z \le h$ it is convenient to rewrite Maxwell equations
in the form of the first order differential equations,
${\partial A}/{\partial y} = M(z) A$, where $M(z)$ is a square matrix
of dimension $8N+4$,
$A^T =(E_y^N\ldots E_y^{-N}, E_x^{N}\ldots E_x^{-N},
H_y^N\ldots H_y^{-N}, H_x^{N}\ldots H_x^{-N})$ and $2N+1$ is the
number of coefficients considered in the
expansion (\ref{eq49}) for $E_y$ and similar expansions for
$E_x,\>H_y$ and $H_x$.

For a rectangular grating the matrix $M(z)$ is constant
(independent of $z$). From Maxwell equations
\begin{eqnarray}
i k_y E_x- \frac{\partial E_y}{\partial x} &=& - i \omega H_z,
\label{eq47}\\
i k_y H_x - \frac{\partial H_y}{\partial x} &=&  i\omega \varepsilon E_z
\nonumber
\end{eqnarray}
\noindent
 we get
 \begin{eqnarray}
 E_z &=& \frac{1}{i\omega \varepsilon}
\left(i k_y H_x - \frac{\partial H_y}{\partial x}\right),
\label{eq48} \\
 H_z &=& -\frac{1}{i\omega }
\left(i k_y E_x - \frac{\partial E_y}{ \partial x}\right).
\nonumber
\end{eqnarray}
\noindent
 Substituting into Eq.~(\ref{eq48}) the expansion (\ref{eq49})
 and analogous expansions for all other components of
 {\boldmath$E$} and {\boldmath$H$}, we arrive at
\begin{eqnarray}
E_z^n &= &\sum_m \Bigl(\frac{1}{i\omega \varepsilon}\Bigr)_{n-m} (i k_y
H_x^m -
i\alpha_m H_y^m),
\nonumber \\
H_z^n &=& -\frac{1}{i\omega} (i k_y E_x^n - i \alpha_n E_y^n),
\label{eq50}
\end{eqnarray}
\noindent
where $(\Theta)_{n-m}$ is the Toeplitz matrix.

The following equations follow from the remaining
four Maxwell equations:
\begin{eqnarray}
\frac{dE_y^n}{dz} &=& i k_y E_z^n - i\omega H_x^n,
\nonumber \\
\frac{dE_x^n}{dz} &=& i\alpha_n E_z^n + i\omega H_y^n,
\label{eq51}\\
\frac{dH_y^n}{dz} &=& i k_y H_z^n +
\sum_m (i\omega\varepsilon)_{n-m} E_x^m,
\nonumber \\
\frac{dH_x^n}{dz} &=& i\alpha_n H_z^n -
\sum_m (i\omega\varepsilon)_{n-m} E_y^m.
\nonumber
\end{eqnarray}
\noindent
One can substitute Eq.~(\ref{eq50})
into Eq.~(\ref{eq51}) and obtain a system of the
first order differential equations for the Fourier components of the
electromagnetic field $E_y^n, E_x^n, H_y^n, H_x^n$ in the region
$0\le z \le h$.

Now we have to determine the
Rayleigh coefficients $R_{np}^{(e)},
R_{np}^{(h)}$ for the specific periodic geometry profile. One can
determine these  coefficients by matching the solution
of equations inside the corrugation region $0\le z \le h$ with
Rayleigh expansions (\ref{eq43}) at $z=h$ and expansions
(\ref{eq45}) at $z=0$. This can be done by imposing the
continuity conditions on each Fourier component of the fields $E_y,
E_x, H_y, H_x$ at $z=0$ and $z=h$.

There is no separation of the TE and TM modes
for the general case. That
is why the reflection matrix $R_{\rm bot}^{(1)}$ for the reflection from
the bottom grating can be defined as follows:
\begin{equation}
R_{\rm bot}^{(1)}(\mbox{\boldmath$k$}_{\bot},\omega) =
\begin{pmatrix} R_{n_1
q_1}^{(e)}(I_p^{(e)}=\delta_{p q_1}, I_p^{(h)}=0  ) \quad
&R_{n_2 q_2}^{(e)}(I_p^{(e)}=0, I_p^{(h)}=\delta_{p q_2} )   \\
R_{n_3 q_3}^{(h)}(I_p^{(e)}=\delta_{p q_3}, I_p^{(h)}=0  ) \quad
&R_{n_4 q_4}^{(h)}(I_p^{(e)}=0, I_p^{(h)}=\delta_{p q_4} )
\end{pmatrix} . \label{52}
\end{equation}

To obtain the Casimir energy we need to determine the
eigenfrequencies of all the normal modes of the electromagnetic field
between the two periodic gratings. These eigenfrequencies can be summed
up by making use of an argument principle, which states:
\begin{equation}
\frac{1}{2\pi i} \oint \phi(\omega) \frac{d}{d\omega} \ln f(\omega)
d\omega = \sum \phi (\omega_0) -\sum \phi(\omega_\infty) .
\label{eq53}
\end{equation}
\noindent
Here, $\omega_0$ are zeroes, $\omega_\infty$ are poles of the
function $f(\omega)$ inside the contour of integration, and the degenerate
eigenvalues are summed over according to their multiplicities. For
the Casimir energy we have $\phi(\omega) = \hbar\omega/2$. The
equation for eigenfrequencies of the corresponding problem of
classical electrodynamics is $f(\omega)=0$.

Consider first the plate-plate geometry when the two dielectric parallel
slabs (slab $1$: $z<0$, slab $2$: $z>L$) are separated by a vacuum
gap ($0<z<L$). In this case  the TE and TM modes are not coupled.
The equation for the TE eigenfrequencies is:
\begin{equation}
f(\omega)= 1-r_{\rm TE, bot}^{(1)}
(\mbox{\boldmath$k$}_{\bot},\omega)
r_{\rm TE, top}^{(2)} (\mbox{\boldmath$k$}_{\bot},\omega) = 0 .
\label{eq54}
\end{equation}
Here $r_{\rm TE, bot}^{(1)}(\mbox{\boldmath$k$}_{\bot},\omega)$
is  the reflection coefficient  of a
plane wave moving down which reflects on the dielectric surface of
the slab
$1$ at $z=0$, while
$r_{\rm TE, top}^{(2)} (\mbox{\boldmath$k$}_{\bot},\omega)$ is the reflection
coefficient of an upward moving
plane wave which reflects on the dielectric
surface of the slab $2$ at $z=L$, and
$\mbox{\boldmath$k$}_{\bot}=(k_x,k_y)$.
One can deduce from the Maxwell
equations that
$r_{\rm TE, top}^{(2)}(\mbox{\boldmath$k$}_{\bot},\omega)=
r_{\rm TE, bot}^{(2)}(\mbox{\boldmath$k$}_{\bot},\omega)\exp(2ik_z L)$
[here, $r_{\rm TE, bot}^{(2)}(\mbox{\boldmath$k$}_{\bot},\omega)$
is the reflection coefficient
of a downward moving TE plane wave which reflects from the dielectric slab
$2$ temporarily located at the position of the slab $1$, i.e. at
$z<0$]. From (\ref{eq54}) and the analogous equation for  the TM modes
one immediately obtains the Lifshitz formula by making use of the
argument principle (\ref{eq53}).

For two periodic dielectrics or gratings separated by a vacuum gap
one has to consider the reflection of downward and upward
moving waves from the
unit cell $0<k_x<2\pi/\Lambda$. Imagine that we removed the top grating
in the system. Then the reflection matrix of the downward
moving wave is defined
as $R_{\rm bot}^{(1)}$. Imagine now that we remove the bottom grating
in the system. Then we denote the reflection matrix of the upward
moving wave as
$R_{\rm top}^{(2)}$. The reflection matrices
$R_{\rm bot}^{(1)},\, R_{\rm top}^{(2)}$
depend on the wave vectors of the incident waves, parameters of the
gratings and the mutual location of the gratings. The equation for
normal modes states:
\begin{equation}
R_{\rm bot}^{(1)} (\mbox{\boldmath$k$}_{\bot},\omega_i)
R_{\rm top}^{(2)}(\mbox{\boldmath$k$}_{\bot},\omega_i,L,x_0) \psi_i = \psi_i,
\label{eq55}
\end{equation}
\noindent
where $\psi_i$ is an eigenvector describing the normal mode with the
frequency $\omega_i$.
 Instead of
Eq.~(\ref{eq54}) one obtains:
\begin{equation}
\det\left[I - R_{\rm bot}^{(1)} (\mbox{\boldmath$k$}_{\bot},\omega)
R_{\rm top}^{(2)}(\mbox{\boldmath$k$}_{\bot},\omega,L,x_0)\right]
 = 0 .\label{eq56}
\end{equation}
For every $k_x,\, k_y$ the solution of (\ref{eq56}) yields  possible
eigenfrequencies $\omega_i$ for the solutions of Maxwell equations
that should be substituted into the definition of the Casimir energy
$E = \sum_i \hbar \omega_i/2$.

Suppose that the reflection matrix $R_{\rm bot}^{(2)}$ for the reflection
from the fictitious (imaginary) grating located as in
Fig.~\ref{grat2} is known in the coordinates $(x,z)$. Performing a
change of coordinates $z= - z_1 +L$, $x=x_1 - x_0 \quad (x_0 <
\Lambda)$ in (\ref{eq43}), it is possible to obtain the
matrix $R_{\rm top}^{(2)}$ for the reflection of upward waves from a grating
with the same profile turned upside-down and  displaced from the lower
grating by $\Delta x=x_0,\, \Delta z=L$ (see Fig.~\ref{grat3}). It
follows that
\begin{equation}
R_{\rm top}^{(2)}(\mbox{\boldmath$k$}_{\bot},i\xi,L,x_0)=
Q^*(x_0) K(\mbox{\boldmath$k$}_{\bot},i\xi,L)
R_{\rm bot}^{(2)}(\mbox{\boldmath$k$}_{\bot},i\xi)
K(\mbox{\boldmath$k$}_{\bot},i\xi,L) Q(x_0) ,
\label{eq57}
\end{equation}
where $R_{\rm bot}^{(2)}(\mbox{\boldmath$k$}_{\bot},i\xi)$
is the reflection matrix of the downward moving
waves from the grating in the system of coordinates $(x,z)$ depicted
in Fig.~\ref{grat2}. Here, $K(\mbox{\boldmath$k$}_{\bot},i\xi,L)$
is the diagonal $2(2N+1)$
matrix of the form:
\begin{equation}
K(\mbox{\boldmath$k$}_{\bot},i\xi,L) = \begin{pmatrix} G_1 &  0  \\
0  &  G_1
\end{pmatrix},  \label{eq58}
\end{equation}
\noindent
with the matrix elements
$\exp[-L\sqrt{\xi^2+k_y^2+(k_x+2\pi m/\Lambda)^2}]$
$(m=-N,\,\ldots,\, N)$ on the main diagonal of the
matrix $G_1$. The lateral translation $2(2N+1)$ diagonal matrix $Q$
is defined as follows:
\begin{equation}
Q(x_0) = \begin{pmatrix}
G_2 &  0  \\
0  &  G_2
\end{pmatrix} ,
\label{eq59}
\end{equation}
\noindent
with matrix elements $\exp(2\pi i m x_0 / \Lambda)$
 on the main diagonal of the matrix $G_2$,
where $m=-N,\,\ldots,\, N$.

 The summation over the eigenfrequencies is
performed by making use of the argument principle (\ref{eq53}), which
yields the Casimir energy of two parallel gratings on a unit cell of
a period $\Lambda$ and unit length in the $y$ direction:
\begin{equation}
E^{\rm C}(L,x_0) = \frac{\hbar c \: \Lambda}{(2 \pi)^3}
\int_0^{\infty} \!\!\!\!d \xi
\int_{-\infty}^{\infty} \!\!\!\!d k_y   \int_0^{2\pi/\Lambda}\!\!\! d k_x
\ln {\rm det}
\Bigl[I - R_{\rm bot}^{(1)} (\mbox{\boldmath$k$}_{\bot},i\xi)
R_{\rm top}^{(2)}(\mbox{\boldmath$k$}_{\bot},i \xi,L,x_0)
\Bigr] . \label{eq60}
\end{equation}
\noindent
Here, as was defined in Sec.~II,
$\varphi = 2\pi x_0/ \Lambda $ and  $x_0$ is a lateral
displacement of the two gratings.
This is an exact expression
valid at zero temperature for two arbitrary parallel gratings with
equal periods $\Lambda$ separated by a vacuum gap.

The Casimir free energy on a unit surface $\mathcal{F}^{\rm C}$
for the system of two
gratings can be written as follows:
\begin{equation}
\mathcal{F}^{\rm C}(L,\varphi) = \frac{k_B T}{\pi^2}
\sum_{l=0}^{\infty}{\vphantom{\sum}}^{\prime}
\int_0^{\infty} \!\!\!dk_y \int_0^{\pi/\Lambda} \!\!\!dk_x \ln {\rm det}
\Bigl[I - R_{\rm bot}^{(1)}(\mbox{\boldmath$k$}_{\bot}, i\xi_l)
R_{\rm top}^{(2)}(\mbox{\boldmath$k$}_{\bot},i \xi_l,L,\varphi) \Bigr].
\label{eq61}
\end{equation}
\noindent
 This formula is valid for an arbitrary profile and an arbitrary
dielectric permittivity of each grating.

When one grating has a curvature of the sphere of the radius
$R\gg \Lambda$, the lateral Casimir force in this system
$F_{\rm lat}^{\rm C}$ can be
obtained by combining the use of the PFA for a sphere-plate configuration
and the exact formula for the free energy of the two
gratings (\ref{eq61}):
\begin{equation}
F_{\rm lat}^{\rm C}(L,\varphi) =  2\pi R  \frac{2\pi}{\Lambda}
\int_{L}^{\infty}\!\!\!
dz^{\prime}\,
\frac{\partial \mathcal{F}^{\rm C}(z^{\prime},\varphi)}{\partial \varphi }.
\label{eq62}
\end{equation}

For the sinusoidally corrugated sphere and plate used in our experiment
$L=a+A_1+A_2$ and the lateral Casimir force is given by
\begin{eqnarray}
F_{\rm lat}^{\rm C}(a,\varphi)& = &\frac{4k_B T R}{\Lambda}
\int_{a}^{\infty}\!\!\! dz^{\prime}
{\sum_{l=0}^{\infty}}{\vphantom{\sum}}^{\prime}
\frac{\partial}{\partial \varphi}
\int_{0}^{\infty}\!\!\! dk_y \int_{0}^{\pi/\Lambda}\!\!\! dk_x
\nonumber\\
&&\times
\ln {\rm det}
\Bigl[I - R_{\rm bot}^{(1)} (\mbox{\boldmath$k$}_{\bot}, i\xi_l)
R_{\rm top}^{(2)}(\mbox{\boldmath$k$}_{\bot},i\xi_l,
z^{\prime}+A_1+A_2,\varphi)
\Bigr] .
\label{eq63}
\end{eqnarray}

Different methods were developed in grating theory starting with
the pioneering work of Rayleigh \cite{71}. For our case of
sinusoidal gratings we used the so-called differential method \cite{72}.
The integration of the first-order ordinary differential equations
(\ref{eq51}) with the dielectric permittivity of the generalized
plasma-like model (\ref{eq36})
in the corrugation region of each grating was based on an explicit
Runge-Kutta $(4,5)$ formula, the Dormand-Prince pair, and was
performed using the Matlab package.
Then the reflection matrices $R_{\rm bot}^{(1)}$ and
$R_{\rm top}^{(2)}$ were computed. The lateral Casimir force
$F_{\rm lat}^{\rm C}$ was computed using Eq.~(\ref{eq63}) for the
experimental parameters of the first and second set of measurement.
Every Matsubara term in Eq.~(\ref{eq63})
was evaluated with a precision of 1.5\% which is also the computational
precision of the obtained $F_{\rm lat}^{\rm C}$.

In Fig.~\ref{phaseN}(a-c) the exact computational results for the lateral
Casimir force as a function of the phase shift between the sinusoidally
corrugations on both surfaces are shown by the solid lines
at separations $a=124.7$, 128.6, and 149.8\,nm, respectively (the
the first set of measurements with corrugation amplitude on the sphere
of 13.7\,nm).
No fitting parameters are used in the comparison
 As can be seen in Fig.~\ref{phaseN}(a-c),
the solid lines are in a very good agreement with the experimental data
shown as dots. These lines clearly demonstrate deviations from the
sinusoidal behavior which decreases with the increase of separation.
Thus, both the experimental data and the exact theory confirm the
prediction of Ref.~\cite{38} made using the PFA approach that the lateral
Casimir force is asymmetric.

In Fig.~\ref{ampN} the exact computational results for the
$\max|F_{\rm lat}^{\rm C}|$ are plotted as a band between the solid lines
versus the separation $a$ between the mean levels of corrugations.
The width of the band takes into account the computational errors
equal to 1.5\% and the
correction to Eq.~(\ref{eq63}) due to surface roughness.
The lower solid line is obtained as the computed $\max|F_{\rm lat}^{\rm C}|$
minus $0.015\max|F_{\rm lat}^{\rm C}|$. The upper solid line
represents $(1+0.015+\eta_{\rm corr})\max|F_{\rm lat}^{\rm C}|$, where
the correction factor  $\eta_{\rm corr}$ for the first set of
measurement was found in Sec.~V. As can be seen in Fig.~\ref{ampN},
the exact theory is in a very good agreement with the measurement data
although it deviates significantly from the theoretical band between the
two dashed lines computed using the PFA approach.
Again no fitting parameters are used in the theory.
This allows one to conclude
that with the grating period of $\Lambda=574.7\,$nm the diffraction-type
effects on the lateral Casimir force have been reliably demonstrated both
experimentally and theoretically.

The results of similar computations using Eq.~(\ref{eq63}) are presented
by the solid lines in Fig.~\ref{phaseM}(a-c) where the lateral Casimir
force is plotted as a function of the phase shift for the second
set
of our mesurements at $a=134$, 156.5, and 179\,nm, respectively.
As before no fitting parameters are used.
In spite of the fact that the experimental data of this set of
measurements with deeper corrugations on the sphere are more noisy,
the exact theory is clearly consistent with data and confirms the
asymmetric (nonsinusoidal) character of the lateral Casimir force.
In Fig.~\ref{ampM} the exact computational results for the
$\max|F_{\rm lat}^{\rm C}|$ in the second set of our measurements versus
separation are presented as the band between the two solid lines.
The width of this band takes into account both the computational errors
and the surface roughness. It was found using the same
procedure, as for the first
set of  measurements, with the replacement of $\eta_{\rm corr}$ by
$\tilde{\eta}_{\rm corr}$ (see Sec.~V).
As in previous comparisons, no fitting parameters are used.
It can be seen in Fig.~\ref{ampM}
that in spite of the larger noise in the second set of measurements,
the data are consistent with the exact theory. The deviations of both
the exact theory and the measurement data from the theoretical prediction
of the PFA approach are even more pronounced than in the first set of
measurements. Thus, the second set of our measurements confirms the
observation of the diffraction effects in the lateral Casimir force.

\section{Conclusions and discussion}

In the foregoing we have presented the results of an experimental and
theoretical investigation of the lateral Casimir force which arises
between Au-coated aligned sinusoidally corrugated surfaces of a sphere
and a plate spaced in close proximity to each other.
The most distinctive feature of this experiment in comparison with the
one performed previously \cite{37,38} is the use of the grating with
much smaller corrugation period which permitted to enter the region of
parameters outside the applicability of the PFA and observe the
nontrivial diffraction-type effects. Another distinctive
feature of this experiment is the use of much deeper corrugations on
both  the plate and the sphere. This allowed us to observe an asymmetry
of the lateral Casimir force (i.e., the deviation of its profile from
perfect sine) which was predicted \cite{38} in the framework of the PFA
approach, but not observed up to date due to the use of small
corrugation amplitudes. In this paper we have presented details of the
experimental setup, the original procedure allowing to imprint relatively
deep corrugations using a metallized
template grating on a sphere of about
$100\,\mu$m radius, and the process of electrostatic calibrations giving
the possibility to measure small forces and at short absolute separations
with high precision. The dependence of the lateral Casimir force on the
phase shift between the corrugations on the
 sphere and the plate was measured at many
different separation distances in the two sets of measurements with
different amplitudes of corrugations on the sphere. The maximum value
of the lateral force as a function of separation was also investigated
in the two sets of measurements. Both the random and systematic errors of
the measured forces and separations were found and the total experimental
errors were calculated at a 95\% confidence level.

The experimental data were compared with an approximate theory using the PFA
and with the exact theory based on the scattering approach. Note that
exact calculation methods allowing the evaluation of the Casimir force
between bodies of arbitrary shape and made of any real material
have been actively investigated since 2006
\cite{35,36,61,62,63,64,65,66,66a,67,68,69,70}.
However, until the present paper and the rapid communication \cite{41}
there was no detailed comparison of exact theoretical results
beyond the Lifshitz theory taking into
account such experimental conditions as real material properties of the
test bodies, surface roughness and nonzero temperature with
measurement data of any specific experiment. Using the exact
theory, the computed dependences of the lateral Casimir force on the
phase shift and of the maximum magnitude of the
lateral force as a function of separation
were found to be in agreement with the experimental data.
Regarding the PFA approach, the computational results were found to be excluded
by the data. This provides the quantitative confirmation for the fact that
this experiment was performed outside the applicability region of the PFA
and marks the beginning of investigations of diffraction effects in
the phenomenon of the lateral Casimir force.

As was already noted in the Introduction,  the lateral Casimir
force might find applications in nanotechnology. Specifically,
it was proposed to use this effect for the frictionless transmission of
lateral motion by means of a nanoscale rack and pinion without
intermeshing cogs \cite{28,29} or in a ratchet with asymmetric corrugations
driven by the Casimir force \cite{30}. The above experimental and
theoretical results bring us closer to the realization of such
kinds of
micromachines. In future it is planned to perform similar experiments with
corrugations of more complicated shape which provides an opportunity
to modulate the respective lateral Casimir force.

\section*{Acknowledgment}
This work was supported by the NSF Grant No.\ PHY0653657
(measurement of the lateral Casimir force) and DOE Grant
No.\ DE-FG02-04ER46131 (calculation of the electric and Casimir
forces).
V.N.M.\ and U.M.\ were also supported by the DARPA Grant
N66001-09-1-2069.
G.L.K.\ and V.M.M.\ were partially supported by the DFG Grant
GE 696/10-1.
V.N.M.\  was partially supported by the
Grants RNP 2.1.1/1575 and RFBR 07-01-00692-a.
The authors are grateful to A.\ V.\ Boukhanovsky,
I.\ S.\ Zudina and to the Institute of Computer Technologies
of St.Petersburg State University of Information
Technologies, Mechanics and Optics where the
numerical computations were performed.


\begin{figure}[h]
\vspace*{13.5cm} 
\caption{\label{setup}Schematic of the experimental setup
(see text for further details).
Insertion shows the obstructed side of the sphere with
imprinted corrugations.}
\end{figure}
\begin{figure*}[h]
\centerline{
} \vspace*{13.5cm} \caption{\label{rghPl}(Color online)
(a) An AFM scan of the grating surface showing the sinusoidal
corrugations covered with stochastic roughness.
(b) A typical section of the grating surface along a
$y=\rm{const}$ plane. The solid line shows a sine function
obtained from the fit. }
\end{figure*}

\begin{figure*}[h]
\centerline{
}
\vspace*{13.5cm} \caption{\label{rghSph}(Color online)
(a) An AFM scan of the  surface of the sphere used in the first set
of measurements showing the imprinted sinusoidal
corrugations covered with stochastic roughness.
(b) A typical section of the grating surface along a
$y=\rm{const}$ plane. The solid line shows a sine function
obtained from the fit.
 }
\end{figure*}

\begin{figure*}[h]
\centerline{
} \vspace*{13.5cm} \caption{\label{sph1}(Color online) The imprinted
corrugations on the sphere used in the first set
of measurements.
The lighter area shows higher points and hence
demonstrates the sphericity of the imprinted surface.}
\end{figure*}
\begin{figure*}[h]
\centerline{
} \vspace*{13.5cm} \caption{\label{sph2}(Color online) The imprinted
corrugations on the sphere  used in the second set
of measurements.
The lighter area shows higher points and hence
demonstrates the sphericity of the imprinted surface.}
\end{figure*}
\begin{figure*}[h]
\centerline{
} \vspace*{13.3cm} \caption{\label{elFc}
The deflection signal due to the lateral electric force versus the
phase shift (a) before and (b) after the correction to the tilt
between the grating and the $x$-axis was introduced.}
\end{figure*}
\begin{figure*}[h]
\centerline{
} \vspace*{13.5cm} \caption{\label{phaseN}(Color online) The
experiment (dots) and the exact theory (solid
lines) for the lateral Casimir force versus the
lateral displacement normalized for the
corrugation period at the separations (a) 124.7\,nm, (b) 128.6\,nm, and
(c) 149.8\,nm in the first set of measurements with corrugation
amplitude on the sphere of 13.7\,nm. No fitting parameters are used.
}
\end{figure*}
\begin{figure*}[h]
\centerline{
} \vspace*{13.5cm} \caption{\label{ampN} The
experimental data (crosses) and theoretical values
computed using the exact theory
(the band between the solid lines) and using the proximity force
approximation (the band between the dashed lines)
for the maximum magnitudes of the
lateral Casimir force versus  separation in the first set of measurements.
No fitting parameters are used.}
\end{figure*}
\begin{figure*}[h]
\centerline{
} \vspace*{13.5cm} \caption{\label{phaseM}(Color online) The
experiment (dots) and  the exact theory (solid lines) for the
lateral Casimir force versus
the lateral displacement normalized for the
corrugation period at the separations (a) 134\,nm, (b) 156.9\,nm, and
(c) 179\,nm in the second set of measurements with corrugation
amplitude on the sphere of 25.5\,nm.
No fitting parameters are used.}
\end{figure*}
\begin{figure*}[h]
\centerline{
} \vspace*{13.5cm} \caption{\label{ampM} The
experimental data (crosses) and theoretical values
computed using the exact theory
(the band between the solid lines) and using the proximity force
approximation (the band between the dashed lines)
for the maximum magnitudes of the
lateral Casimir force versus  separation in the
second set of measurements.
No fitting parameters are used.}
\end{figure*}
\begin{figure*}[h]
\centerline{
} \vspace*{13.5cm} \caption{\label{grat1}(Color online) Two
longitudinal gratings of different shape
and same period $\Lambda$.}
\end{figure*}
\begin{figure*}[h]
\centerline{
} \vspace*{13.5cm} \caption{\label{grat2}(Color online) The
fictitious longitudinal grating for which one evaluates
the reflection matrix $R_{\rm bot}^{(2)}$
(see text for further discussion). }
\end{figure*}
\begin{figure*}[h]
\centerline{
} \vspace*{13.5cm} \caption{\label{grat3}(Color online) The
top longitudinal grating in Fig.~\ref{grat1} for which one evaluates
the reflection matrix $R_{\rm top}^{(2)}$.
The normal and lateral shifts with respect to the fictitious grating
shown in Fig.~\ref{grat2} are denoted by $L$ and $x_0$,
respectively. }
\end{figure*}
\begingroup
\squeezetable
\begin{table}
\caption{\label{tab1} The mean values (column 2),
the variances of the mean (column 3),
the systematic errors (column 4),
and the total experimental errors at a 95\% confidence
level (column 5) of the measured maximum magnitudes
of the lateral Casimir force $f\equiv\max|F_{\rm lat}^{\rm C}|$
 at different separations (column 1) for the first set of
 measurements.}
\begin{ruledtabular}
\begin{tabular}{ccccc}
$a$ & $\bar{f}$ & $s_{\bar{f}}$ & $\Delta^{\!\rm syst}f$ &
$\Delta^{\!\rm tot}f$  \\
(nm) & (pN)& (pN) & (pN) & (pN) \\
\hline
121.1 & 49.3 &3.6 & 2.4 & 11.1 \\
124.7 & 36.4 & 1.3 & 1.7 & 4.7 \\
128.6 & 27.8 & 2.4 & 1.3 & 7.1 \\
137.3 & 17.6 & 0.72 & 0.86 & 2.5 \\
149.8 & 9.50 & 0.56 & 0.48 & 1.8 \\
162.6 & 5.76 & 0.23 & 0.30 & 0.8 \\
175.3 & 3.71 & 0.37 & 0.20 & 1.1 \\
188.1 & 2.33 & 0.40 & 0.13 & 1.1
\end{tabular}
\end{ruledtabular}
\end{table}
\endgroup
\begingroup
\squeezetable
\begin{table}
\caption{\label{tab2} The mean values (column 2),
the variances of the mean (column 3),
the systematic errors (column 4),
and the total experimental errors at a 95\% confidence
level (column 5) of the measured maximum magnitudes
of the lateral Casimir force $f\equiv\max|F_{\rm lat}^{\rm C}|$
 at different separations (column 1) for the second set of
 measurements.}
\begin{ruledtabular}
\begin{tabular}{ccccc}
$a$ & $\bar{f}$ & $s_{\bar{f}}$ & $\Delta^{\!\rm syst}f$ &
$\Delta^{\!\rm tot}f$  \\
(nm) & (pN)& (pN) & (pN) & (pN) \\
\hline
133.9 & 52.5 & 4.5 & 2.5 & 12.0 \\
145.2 & 24.1 & 1.35 & 1.2 & 3.9 \\
156.5 & 13.6 & 0.60 & 0.67 & 1.9 \\
179.0 & 6.62 & 0.26 & 0.35 & 0.86 \\
201.6 & 3.49 & 0.24 & 0.20 & 0.69\\
224.1 & 2.37 & 0.31 & 0.15 & 0.81 \\
246.6 & 1.48 & 0.25 & 0.10 & 0.63
\end{tabular}
\end{ruledtabular}
\end{table}
\endgroup
\end{document}